\documentclass[fleqn,12pt]{article}
\usepackage{amsmath}
\usepackage{amsfonts,epsfig,latexsym}
\usepackage[vcentermath,enableskew]{youngtab}
\usepackage[nosort]{cite}
\def\baselinestretch{1.1}
\newdimen\squaresize \squaresize=12pt
\newdimen\thickness \thickness=0.7pt

\def\square#1{\hbox{\vrule width \thickness
   \vbox to \squaresize{\hrule height \thickness\vss
      \hbox to \squaresize{\hss#1\hss}
   \vss\hrule height\thickness}
\unskip\vrule width \thickness} \kern-\thickness}

\def\cut#1{\hbox{\vrule width-1 \thickness
   \vbox to \squaresize{\hrule height-1 \thickness\vss
      \hbox to \squaresize{\hss#1\hss}
   \vss\hrule height-1\thickness}
\unskip\vrule width +4 \thickness} \kern-\thickness}

\def\vsquare#1{\vbox{\square{$#1$}}\kern-\thickness}

\newcommand{\ft}[2]{{\textstyle\frac{#1}{#2}}}
\def\tilde{\widetilde}
\def\1bar{1\hskip -.275cm -}
\def\2bar{2\hskip -.275cm -}
\def\3bar{3\hskip -.275cm -}

\newsavebox{\uuunit}

\newcommand{\pls}{\!+\!}

\newcommand{\mis}{\!-\!}

\newcommand{\mathon}{\mathversion{bold}}
\newcommand{\mathoff}{\mathversion{normal}}
\newcommand{\la}{\label}

\newcommand{\ie}{{\it i.e.~}}
\newcommand{\eg}{{\it e.g.~}}
\newcommand{\ap}{\alpha^{\prime}}

\newcommand{\be}{\begin{equation}} \newcommand{\ee}{\end{equation}}
\newcommand{\bea}{\begin{eqnarray}} \newcommand{\eea}{\end{eqnarray}}
\newcommand{\ben}{\begin{displaymath}}
\newcommand{\een}{\end{displaymath}}
 
\newcommand{\nn}{\nonumber} \newcommand{\non}{\nonumber\\}

\newcommand{\spartial}{\partial\hspace{-2 mm}\slash}

\newcommand{\jb}{\bar{\jmath}}

\makeatletter
\let\old@makecaption=\@makecaption
\def\@makecaption{\small\old@makecaption}
\makeatother

\makeatletter \@addtoreset{equation}{section} \makeatother

\makeatletter
\let\old@startsection=\@startsection
\renewcommand{\@startsection}[6]{\old@startsection{#1}{#2}{#3}{#4}{#5}{#6\mathversion{bold}}}
\makeatother

\let\oldPhi=\Phi
\let\oldPsi=\Psi
\let\oldGamma=\Gamma
\let\oldSigma=\Sigma
\renewcommand{\Phi}{\mathnormal{\oldPhi}}
\renewcommand{\Psi}{\mathnormal{\oldPsi}}
\renewcommand{\Gamma}{\mathnormal{\oldGamma}}
\renewcommand{\Sigma}{\mathnormal{\oldSigma}}

\newcommand{\hypref}[2]{\ifx\href\asklfhas #2\else\href{#1}{#2}\fi}

\newcommand{\sfrac}[2]{{\textstyle\frac{#1}{#2}}}

\newcommand{\alg}[1]{\mathfrak{#1}}

\newcommand{\alSU}{\alg{su}}

\newcommand{\alSO}{\alg{so}}

\newcommand{\alPSU}{\alg{psu}}
\newcommand{\alhs}{\alg{hs}}
\newcommand{\mult}{\mathcal{V}}

\newcommand{\superN}{\mathcal{N}}
\newcommand{\gym}{g_{\scriptscriptstyle\mathrm{YM}}}

\newcommand{\bigbrk}[1]{\bigl(#1\bigr)}

\newcommand{\nln}{\nonumber\\}
\newcommand{\nl}{\nonumber\\&&\mathord{}}

\newcommand{\earel}[1]{\mathrel{}&#1&\mathrel{}}
\newcommand{\eq}{\earel{=}}
\newenvironment{myeqnarray}{\arraycolsep0pt\begin{eqnarray}}{\end{eqnarray}\ignorespacesafterend}
\newenvironment{myeqnarray*}{\arraycolsep0pt\begin{eqnarray*}}{\end{eqnarray*}\ignorespacesafterend}

\def\[{\begin{equation}}
\def\]{\end{equation}}
\def\<{\begin{myeqnarray}}
\def\>{\end{myeqnarray}}

\begin{document}

\thispagestyle{empty}

\begin{center}

{\small\ttfamily ROM2F/04/27\hspace*{0.8cm} }

\end{center}


\begin{center}

\renewcommand{\thefootnote}{\fnsymbol{footnote}}

{\mathon\bf\Large  Higher spins and stringy 
$AdS_5\times S^5$\footnote{Lecture delivered at the RTN Workshop ``The quantum
structure of spacetime and the geometric nature of fundamental
interactions'', and EXT Workshop ``Fundamental Interactions and the
Structure of Spacetime'' in Kolymbari, Crete, 5-10 September 2004.}
\par \mathoff}%
\bigskip\bigskip

\textbf{M.~Bianchi}

\vspace{.3cm}

\textit{Dipartimento di Fisica and INFN \\
Universit\`a di Roma ``Tor Vergata''\\
00133 Rome, Italy}\\
{\texttt{Massimo.Bianchi@roma2.infn.it}}
\setcounter{footnote}{0}

\end{center}

\begin{abstract}

In this lecture I review recent work done in collaboration with N. Beisert, J.
F. Morales and H. Samtleben \cite{Bianchi:2003wx, Beisert:2003te, 
Beisert:2004di}\footnote{For a concise summary see \cite{Bianchi:2004talk}.}. 
After a notational flash on the AdS/CFT correspondence, I will discuss higher 
spin (HS) symmetry enhancement at small radius and how this is holographically 
captured by free N=4 SYM theory. I will then derive the spectrum 
of perturbative superstring excitations on AdS in this particular limit and 
successfully compare it with
the spectrum of single-trace operators in free ${\cal N}=4$ SYM at
large N, obtained by means of Polya(kov)'s counting. 
Decomposing the spectrum into HS multiplets allows one to
precisely identify the `massless' HS doubleton and the lower spin Goldstone 
multiplets which participate in the pantagruelic Higgs mechanism, termed
{\it ``La Grande Bouffe''}. After recalling some basic features of Vasiliev's 
formulation of HS gauge theories, I will eventually sketch how to 
describe mass generation in the AdS bulk \`a la S\"uckelberg and 
its holographic implications such as the emergence of anomalous dimensions in 
the boundary  ${\cal N}=4$ SYM theory.  

\end{abstract}

\def\baselinestretch{1.1}

{}{}{}{}{}{}{}{}{}{}{}{}{}{}{}{}{}{}{}{}{}{}{}{}{}{}

\section{Introduction and Summary}

The plan of the lecture is as follows.

I will begin with a flash on the AdS/CFT  correspondence with the
purpose of establishing the notation and recalling how semiclassical string
solitons with large spin (S) and/or R-charge (J) in the bulk can
be associated to gauge-invariant composite operators on the
boundary, for reviews see \eg
\cite{Aharony:1999ti, Bianchi:2000vh, D'Hoker:2002aw, Tseytlin:2003ii} and 
references therein.

I will then pass to consider Higher Spin (HS) Symmetry enhancement
at vanishing coupling on the boundary and discuss how this regime
should be captured by the extreme stringy regime of very `small'
radius. 
Exploiting HS enhancement and extrapolating the BMN
formula, I will show how to derive the spectrum of perturbative superstring
excitations on AdS in this particular limit and compare it with
the spectrum of single-trace operators in free ${\cal N}=4$ SYM at
large N, obtained by means of Polya(kov)'s counting. Needless to
say perfect agreement is found up to the level that is possible to
reach by computer-aided human means.

Decomposing the spectrum into HS multiplets allows one to
precisely identify the {`massless'} HS doubleton,
comprising the HS currents on the boundary dual to massless --
in this limit -- gauge fields in the bulk, with the `first Regge
trajectory'. Higher Regge trajectories correspond to {`massive'}
HS $L$-pletons ($YT$-pletons), associated to Yang-tableaux compatible 
with gauge invariance, comprise KK excitations of the doubleton, lower 
spin Goldstone modes as well as genuinely long superconformal multiplets.

Glimpses of {\it ``La Grande Bouffe"}\footnote{Several people
asked me the origin of this terminology: it is the title of a
movie directed by Marco Ferreri, interpreted, among other, by
Marcello Mastroianni and Ugo Tognazzi and presented in 1973 at
{\it Festival du Cinema} in Cannes where it received the
International Critics Award.}, the Pantagruelic Higgs mechanism
whereby HS gauge fields eat lower spin Goldstone fields and become
massive, are presented in the St\"uckelberg formulation that suggests
how the mass shifts should be governed by (broken) HS symmetry. I
will not dwelve too much into the discussion of anomalous
dimensions that emerge in the boundary theory when interactions
are turned on as a result of the resolution of operator mixing.

\section{Notational flash on AdS/CFT}

Maldacena's conjectures triggered an
intense renewal of interest in (super)conformal field theories (SCFT) and
lead to the discovery of previously unknown non-renormalization
properties
\cite{Lee:1998bx, D'Hoker:1999ea, Bianchi:1999ie, Eden:1999kw, D'Hoker:2000dm,
Eden:2000gg}.
In particular the holographic correspondence between ${\cal N}=4$
SYM theory in $d=4$ with $SU(N)$ gauge group and Type IIB
superstring theory on $AdS^5 \times S^5$ with $N$ units of RR
5-form flux has been an inhexaustible source of insights in the
duality between gauge fields and strings. ${\cal N}=4$ SYM theory
is an exactly superconformal field theory at the quantum level.
The elementary field content of the theory consists of one gauge vector
$A_\mu$, ${\bf 4}$ Weyl gaugini $\lambda^A_\alpha$ together with their
 ${\bf 4}^*$ conjugate $\bar\lambda_A^{\dot\alpha}$ and  ${\bf 6}$ real scalars
$\varphi^i$ all in the adjoint representation of the gauge group.
All interactions up to two derivatives are fixed by the choice of
the gauge group, \ie of the structure constants $f_{abc}$, and the
$\beta$-function vanishes both perturbatively and
non-perturbatively\footnote{Due to instanton effects, physical
observables may {\it a priori}  depend on $\theta_{ym}$ since
${\cal N}=4$ SYM has no internal R-symmetry anomaly. Holography
relates the vacuum angle to the axion $\theta_{ym}=2\pi \chi$ and
YM instantons to type IIB D-instantons
\cite{Banks:1998nr, Witten:1998xy, Bianchi:1998nk, Dorey:1998qh,
Dorey:1999pd, Bianchi:2001jg,
Bianchi:2002gz, Green:2002vf, Kovacs:2003rt}.}. The semiclassical
relation between the couplings $g_{ym}^2 = 4\pi g_s$ and the
curvature radius $R^4 = \lambda(\ap)^2$, where $\lambda = g_{ym}^2
N$ is the 't Hooft coupling, suggests that the planar limit,
dominated by amplitudes with the topology of the sphere, is
achieved at large $N$ with fixed small $\lambda$. However one can
trust the low-energy supergravity approximation only at large
$\lambda$ where the curvature is small, a regime that is difficult
to analyze from the SYM perspective except for few observables
protected against quantum corrections by extended superconformal symmetry.
The relavant supergroup, $(P)SU(2,2|4)$, includes the $SU(2,2)\approx
SO(4,2)$ isometry of $AdS_5$ acting as conformal group on the
boundary, and the $SU(4)\approx SO(6)$ isometry of $S^5$, playing
the role of R-symmetry group in ${\cal N}=4$ SYM. Each
gauge-invariant local composite operator ${\cal O}_{\Delta}(x)$ of
scaling dimension $\Delta$ on the (conformal) boundary $\rho
\approx 0$ of $AdS$ is expected to be holographically dual to a bulk field
$\Phi_M(x,\rho)$, associated to a string excitation of AdS mass
$M$. The near boundary behaviour $ \Phi(x,\rho)\approx
\rho^{4-\Delta} j(x) + ... $ is dictated by the boundary source
$j_{4-\Delta}(x)$ that couples to ${\cal O}_{\Delta}(x)$. Linearized
field equations determine the mass-to-dimension relation \be M^2
R^2 = \Delta (\Delta - 4) - \Delta_{u} (\Delta_{u} - 4)\ee where
$\Delta_{u}$ is the lower bound on $\Delta$ imposed by unitarity
of the $(P)SU(2,2|4)$ representation $\Phi_M \approx {\cal O}_{\Delta}$
belongs to.

A very interesting class of operators consists of those in 1/2 BPS
(ultra) short multiplets that correspond to the ${\cal N}=8$
gauged supergravity multiplet and its Kaluza-Klein (KK)
recurrences. Their lowest {\it superprimary} components are scalar 
Chiral Primary
Operators (CPO's) \be {\cal Q}^{(i_1...i_p)|} = Tr ( \varphi^{i_1}
... \varphi^{i_p})\ee of dimension $\Delta = p$ belonging to
the $p$-fold totally symmetric and traceless tensor
representation of the $SO(6)$ R-symmetry with $SU(4)$ Dynkin
labels $[0,p,0]$. CPO's are not only
annihilated by the 16 superconformal charges $S^A_\alpha$,
$\bar{S}_A^{\dot\alpha}$ but also by half of the 16 Poincar\`e
supercharges $Q^A_\alpha$, $\bar{Q}_A^{\dot\alpha}$. Generically,
1/2 BPS multiplets contain $2^8 p^2 (p^2 -1)/12$ components. $p=0$
is the identity while $p=1$ is the {\it singleton} representation
of $PSU(2,2|4)$ corresponding to the elementary SYM fields (8
bosons and 8 fermions) that do not propagate in the bulk. The
${\cal N}=4$ supercurrent multiplet ($p =2$) contains 128 bosonic
and as many fermionic components and includes the conserved
traceless stress tensor ${\cal T}_{\mu\nu}$, the ${\bf 15}$
conserved R-symmetry currents ${\cal J}^{[ij]}_\mu$ and the ${\bf
4^*}$ conserved and $\gamma$-traceless supercurrents
$\Sigma^\alpha_{\mu A}$ as well as their ${\bf 4}$ conjugate
$\bar\Sigma^A_{\mu \dot\alpha}$. Correlation functions of CPO's
 \be G(x_1, ...x_n) = \langle
{\cal Q}_{p_1}(x_1){\cal Q}_{p_2}(x_2)... {\cal Q}_{p_n}(x_n)
\rangle \ee enjoy remarkable (partial) non-renormalization
properties \cite{Lee:1998bx, D'Hoker:1999ea, Bianchi:1999ie, Eden:1999kw,
D'Hoker:2000dm,Eden:2000gg}. Two and three-point functions as well as 
extremal ($p_1 = \sum_{i\neq 1} p_i$) and next-to-extremal 
($p_1 + 2 = \sum_{i\neq
1} p_i$) correlators do not receive any quantum
correction. Near extremal correlators ($p_1 + k = \sum_{i\neq 1}
p_i$, with small $k$), \eg four-point functions of the CPO's
${\cal Q}_{2}$ in the supercurrent multiplet, display some sort of
partial non-renormalization both at weak coupling perturbatively
and non, where field theory methods are reliable, and at strong
coupling where supergravity is reliable.

Massive string excitations correspond to long multiplets with at
least $2^{16}$ components \cite{Andrianopoli:1999vr, Andrianopoli:1998ut}. 
The prototype is the ${\cal N}=4$ Konishi multiplet \cite{Bianchi:2001cm} that 
starts with the scalar singlet operator ${\cal K} = Tr(\varphi_i \varphi^i)$ 
of naive dimension $\Delta_0 = 2$ at vanishing coupling where a unitary bound 
of the semishort kind is saturated and the currents of spin up to 4 at higher 
level are conserved.
When interactions are turned on, all the components of the multiplet acquire 
the same anomalous dimension $\gamma^{\cal K}$, that was computed at one loop 
long ago \cite{Anselmi:1996dd}. The result, $\gamma^{\cal K}_{1-loop} =
+3 \lambda$, has been confirmed and extended to two loops, $\gamma^{\cal
K}_{2-loop} = -3 \lambda^2$, by explicit computations \cite{Bianchi:2001cm,
Bianchi:2000hn} and three
loops, $\gamma^{\cal K}_{3-loop} = 21/4\lambda^3$
\cite{Beisert:2003tq, Beisert:2003ys}, assuming integrability of the 
super spin chain, whose hamiltonian represents the dilatation operator
\cite{Minahan:2002ve, Beisert:2003tq, Beisert:2003yb, 
Beisert:2004hm}.
Instanton corrections are absent to lowest order
\cite{Bianchi:2001cm, Kovacs:2003rt}. At strong 't Hooft coupling
operators dual to string excitations with masses $M^2\approx
1/\ap$ should acquire large anomalous dimensions $\Delta =
\Delta_0 + \gamma \approx MR \approx \lambda^{1/4}$ and decouple
from the operator algebra. Unfortunately, it has been difficult to
test and exploit the correspondence beyond the supergravity
approximation since, despite some progress \cite{Pesando:1998fv,
Pesando:1999wf, Berkovits:1999im, Kallosh:2000bd, Metsaev:2000bj, 
Metsaev:2000yu}

an efficient quantization scheme for type IIB superstring on
$AdS_5\times S^5$ is still lacking. Berkovits's pure spinor
formalism \cite{Berkovits:2002zk} allows one to write covariant
emission vertices for `massless' supergravity fields and their KK
recurrences \cite{Berkovits:2000yr}. Studying the first massive
level, that, as we will argue, corresponds to the Konishi
multiplet and its KK recurrences, should shed new light on the
stringy aspects of the holographic correspondence. 

Alternatively one can consider particular sectors of the spectrum or peculiar
regimes where computations are feasible in both descriptions. On the one hand 
one can study semiclassical string solitons with
large spin or R-charge, whose dynamics can be studied
perturbatively in terms of a reduced coupling $\lambda' =
\lambda/L^2$ where $L$ measures the `length' of the operator /
string \cite{Tseytlin:2003ii}. On the other hand one can try to study the 
string spectrum and interactions in the very stringy regime of small radius 
$R$,
dual to free ${\cal N}=4$ SYM,  where holography predicts Higher
Spin (HS) symmetry enhancement \cite{WittenJHS60, Sundborg:2000wp,
Konstein:2000bi, Sezgin:2001zs}. Before doing that let us briefly
recall some results concerning string solitons in AdS. 
In QCD, processes like Deep Inelastic Scattering can be studied by
means of Operator Products Expansions (OPE's) of local operators
that are dominated by operators with
arbitrary dimension $\Delta$ and spin $s$ but fixed twist $\tau =
\Delta - s$, in particular currents with $\tau = 2 + \gamma$ such as 
\be J^v_{({\mu_1}{\mu_2}... {\mu_s})|} =
Tr (F_{\nu(\mu_1}D_{\mu_2}... D_{\mu_{s-1}}F_{\mu_s)|}{}^\nu )
\quad , \quad J^f_{({\mu_1}{\mu_2}... {\mu_s})|} = \bar\psi_u
\gamma_{(\mu_1}D_{\mu_2}... D_{\mu_{s})|}\psi^u \ee These
mix (beyond one-loop) with one another and, except for
the stress tensor and the conserved vector currents, acquire
anomalous dimensions $\gamma (S)$. At large $S$ the dominant
contribution, including higher loops, goes as $\gamma (S)\approx
\log (S)$. This remarkable gauge theory prediction \cite{Curci:1980uw,
Furmanski:1980cm}
has been
confirmed by string computations \cite{Gubser:2002tv} showing that
the dispersion relation for long folded strings in $AdS_5$ with
large $S$ is of the form \be MR = \Delta  = S + a \sqrt{\lambda}
\log(S) + ... \ee Unfortunately it is difficult at present to
quantitatively study the perturbative contributions in 
$\lambda$ that should reconstruct the $\sqrt{\lambda}$ at large
$\lambda$. It is however reassuring to observe that small strings,
even in $AdS_5$, display the standard relation $M^2 = S/\ap$.

In addition to the above HS currents ${\cal N}=4$ SYM offers the
possibility of studying operators with large R-charge, \ie large
angular momentum $J$ on $S^5$. Decomposing $SO(6)$ under $U(1)_J\times
SU(2)\times SU(2)$, it is easy to see that CPO's of the form
$Tr(Z^J)$ with $\Delta = J$ are 1/2 BPS and protected since there
are no other operators they can possibly mix with (in the planar
limit). All other operators can be built by successively inserting
{\it impurities} with $\Delta > J$. In particular four real
scalars, four of the gaugini and the four derivatives have $\Delta
- J = 1$. Berenstein, Maldacena and Nastase \cite{Berenstein:2002jq}
argued that the sector
of operators with large R-charge is dual to the type IIB
superstring on the maximally supersymmetric pp-wave that emerges
from a Penrose limit of $AdS_5\times S^5$ \cite{Blau:2001ne, Blau:2002dy}. 
Despite the presence of a 
null RR 5-form
flux $F_{+1234} = F_{+5678}=\mu$, superstring fluctuations around
the resulting background can be quantized in the light-cone gauge
\cite{Metsaev:2001bj, Metsaev:2002re} whereby $p^+ = J/(\mu \ap)$ and
the `vacuum' $|p^+\rangle$ corresponds to $Tr(Z^J)$. The
spectrum of the light-cone Hamiltonian \be H_{LC} = p^- = \mu
(\Delta - J) = \mu \sum_n N_n \omega_n \quad , \quad \omega_n =
\sqrt{ 1 + {n^2 \lambda\over J^2}} \quad ,\ee combined with the
level matching condition $\sum_n n N_n=0$, thus gives a prediction for
the anomalous dimensions of so-called BMN operators with large
R-charge. BMN operators with
$\Delta= J+1$ (one impurity) are necessarily superconformal descendants of the
`vacuum' and are thus protected. Operators with two
impurities, say $X$ and $Y$, \be a^X_n a^Y_{-n} |p^+\rangle \quad
\leftrightarrow \quad \sum_k e^{2\pi i k n /J} Tr(X Z^{J-k} Y Z^k)
\ee are in general unprotected, since $\Delta= J + 2\omega_n >
J+2$ for $n\neq 0$. At large but finite $J$ BMN operators form
$(P)SU(2,2|4)$ multiplets \cite{Beisert:2002bb} whose structure
becomes more and more involved with the number of impurities. For
our later purposes it is crucial that any single-trace
operator in ${\cal N}=4$ can be identified with some component of a
BMN multiplet with an arbitrary but finite number of impurities.

\section{Superstring spectrum on stringy $AdS_5\times S^5$}

At vanishing coupling, free ${\cal N}=4$ SYM
exposes HS Symmetry enhancement
\cite{WittenJHS60, Sundborg:2000wp, Konstein:2000bi, Sezgin:2001zs}.
Conformal invariance indeed implies that a spin $s$ current, such
as \be J_{({\mu_1}{\mu_2}... {\mu_s})|} = Tr (\varphi_i
D_{(\mu_1}... D_{\mu_{s})|}\varphi^i) +... \quad , \ee saturating
the unitary bound $\Delta_0 = 2 + s$ be conserved. ${\cal N}=4$
superconformal symmetry implies that twist two operators are
either conserved currents or superpartners thereof
\cite{Dobrev, Dolan:2001tt, Dolan:2002zh}. Once interactions are turned off
long multiplets decompose into (semi)short ones forming the {\it
doubleton} representation of $HS(2,2|4)$, the HS extension of
$(P)SU(2,2|4)$.

The weak coupling regime on the boundary should be holographically
dual to a highly stringy regime in the bulk, where the curvature
radius $R$ is small in string units $R \approx \sqrt{\ap}$ and the
string is nearly tensionless
\cite{Lindstrom:1993yb, Isberg:1993av}. Although, as remarked
above, quantizing the superstring in $AdS_5\times S^5$ is a
difficult and not yet accomplished task, the huge enhancement of
symmetry allows us to determine the superstring spectrum in this
limit and to precisely match it with the ${\cal N}=4$ SYM spectrum.
To this end, it is first convenient to recall the structure of the
type IIB superstring spectrum in flat spacetime.


In the light-cone GS formalism (left-moving) superstring excitations are
obtained by acting on the groundstate $|{\cal Q}\rangle$ with the
the ${\bf 8}_V$ bosonic, $\alpha^I_{-n}$, with $I=1,...8$, and the the ${\bf
8}_S$ fermionic, $S^a_{-n}$, with $a=1,...8$, creation operators. As a result 
of the quantization of the
fermionic zero-modes $S^a_0$ the groundstate $|{\cal Q}\rangle= |I\rangle -
|\dot{a}\rangle$ is $16$-fold degenerate and consists of ${\bf 8}_V$
bosons and ${\bf 8}_C$ fermions. Combining with right-moving modes
with the same chirality projection on the vacuum and imposing 
level-matching $\ell = \sum_n n N^L_n = \sum_n n N^R_n$ one obtains
the complete physical spectrum of 'transverse' single-particle
excitations. At $\ell = 0$ one finds the components of the type
IIB ${\cal N}=(2,0)$ supergravity multiplet: the graviton $G_{(IJ)|}$,
two antisymmetric tensors $B^r_{[IJ]}$, two scalars $\phi^r$ and a
four-index self-dual antisymmetric tensor $A_{IJKL}$, for a total
of $35+ 2\times 28 + 2 + 35= 128$ bosonic d.o.f.; two gravitini
$\Psi_{Ia}^u$ and two dilatini $\Lambda_{\dot{a}}^u$, for a total
of $2\times(56 +8)=128$ fermionic d.o.f..

At higher levels, $\ell\geq 1$, the (chiral) spectrum not
unexpectedly assembles into full representations of the massive
transverse Lorentz group $SO(9)$. Indeed, focussing for simplicity
on the left-moving sector at lowest level one finds \be \ell =1
\quad : \quad (\alpha^I_{-1} - S^a_{-1})(|J\rangle - |\dot{b}\rangle) \ee
yielding \be ({\bf 8}_V - {\bf 8}_S)({\bf 8}_V - {\bf 8}_C) =
({\bf 1}_O + {\bf 28}_O + {\bf 35}_O + {\bf 8}_V + {\bf 56}_V) -
({\bf 8}_C + {\bf 56}_C + {\bf 8}_S + {\bf 56}_S) \ee that can be
reorganized into $SO(9)$ representations \be \ell = 1 \quad : \quad {\bf 44}
+ {\bf 84} - {\bf 128} \ee corresponding to a symmetric
tensor (`spin 2'), a 3-index totally antisymmetric tensor and a
spin-vector (`spin 3/2'). Incidentally this is exactly the
(massless) field content of $D=11$ supergravity.  For $\ell>1$ the
situation is analogous though more involved.

The next task is to decompose the spectrum in ${\cal N}=(2,0)$
supermultiplets, \ie  identify the groundstates  annihilated by
half of the $32= 2\times 16$ supercharges, that play the role of
`lowering' operators. For $\ell=1$ the groundstate cannot be
anything else than an $SO(9)$ singlet $V^{L/R}_{\ell=1} = {\bf
1}$, \ie a scalar, since $2^8\times 2^8 = 2^{16}$ equals the
number of d.o.f. at this level. At higher level the situation is
not so straightforward. It proves convenient to `factor out' --
for both left- and right-movers -- the structure $ {\cal Q}_S
\times {\cal Q}_C = ({\bf 8}_V - {\bf 8}_S)({\bf 8}_V - {\bf
8}_C)$ that corresponds to the action of the 8 `raising'
supercharges. Using this trick one can deduce a recurrence
relation that yields \be V^{L/R}_{\ell=1}= {\bf 1} \quad , \quad
V^{L/R}_{\ell=2} = {\bf 9} \quad , \quad V^{L/R}_{\ell=3} = {\bf
44}- {\bf 16} \quad , \quad ... \ee for the first few levels. In
summary, the Hilbert space of type IIB superstring excitations in
flat space can be written as\be {\cal H}_{flat} = {\cal
H}_{sugra} + {\cal T}_{susy} \sum_\ell V^{L}_{\ell} \times
V^{R}_{\ell} \ee where ${\cal T}_{susy}=  [{\cal Q}_S \times {\cal
Q}_C]_L \times[{\cal Q}_S \times {\cal Q}_C]_R$ represents the
action of the supercharges.

Two amusing features are worth noticing at this point. First, the
maximum spin at level $\ell$ is $s_{_{Max}}=2\ell + 2$. States with
such spin and their superpartners are said to belong to the first
Regge trajectory which is generated by the oscillators with lowest
possible mode number. Second, the partial sums
$\sum_\ell^{1,K} V^{L}_{\ell} \times V^{R}_{\ell}$ form $SO(10)$
multiplets. This can be related to the possibility of
`covariantizing' the massive spectrum of type IIB, which is
identical to the one of type IIA, to $SO(10)$, by lifting it to
$D=11$ \cite{Bars:1995uh, Bars:2004dg}, or to $SO(9,1)$, by introducing the 
worldsheet (super)ghosts.


In order to extrapolate the massive string spectrum from flat
space to $AdS_5\times S^5$ at the HS spin symmetry enhancement
point one can take the following steps. First, decompose
$SO(9)$ into $SO(4)\times SO(5)$, the relevant stability group of
a massive particle; second, identify the KK towers of
spherical harmonics that replace the (internal) momenta; third, assign an AdS 
mass $M$, dual to the boundary scaling
dimension $\Delta$, to each state.

The first step is completely straightforward and determines two of the quantum 
numbers of the relevant $PSU(2,2|4)$
superisometry group, namely the two spins $(j_L, j_R)$ of
$SO(4)\subset SO(4,2)$. In more covariant terms and to linear
order in fluctuations around $AdS_5\times S^5$, diagonalization of
the type IIB field equations should produce a set of uncoupled
free massive equations: \bea &&\left(\nabla_{AdS_5\times
S^5}^2-M_{\Phi}^2 \right) \Phi_{ \{ \mu \} \,\{ i\}}=0 \label{emh}
\eea The collective indices $\{ \mu \} \in {\cal R}_{SO(1,4)}$ and
$\{ i \} \in {\cal R}_{SO(5)}$ label irreducible representations
of the relevant Lorentz group, ${SO(4,1)}\times SO(5)$.  To each
excitation around flat spacetime we associate a tower of KK
recurrences. The spectrum of AdS masses $M_\Phi$ taking into
account the coupling to the curvature and RR 5-form flux should be
determined by requiring consistency, \ie~BRS invariance, of
superstring propagation on $AdS_5\times S^5$. Each ten-dimensional
field $\Phi_{ \{ \mu \}\,\{i\}}$ can be expanded in
$S^5$-spherical harmonics:
\be \Phi_{ \{ \mu \}\,\{i\}}(x,y)=\sum_{[k,p,q]} {\cal
X}^{[kpq]}_{\{\mu \}}(x)\, {\cal Y}^{[kpq]}_{\{i\}}(y) \;,
\label{harm} \ee
with $x, y$ coordinates along $AdS_5$ and $S^5$ respectively.
The sum runs over the set of allowed representations of the $S^5$
isometry group $SO(6)\approx SU(4)$ characterized by their $SU(4)$
Dynkin labels $[k,p,q]$ and the (generalized) spherical harmonics
${\cal Y}^{[kpq]}_{\{i\}}(y)$ are eigenvectors of the $S^5$ Laplacian:
\bea \nabla_{S^5}^2 \, {\cal Y}^{[kpq]}_{\{i\}} &=&
-\left(C_2\left[ SO(6)\right]-C_2\left[ SO(5)\right]\right)
\,{\cal Y}^{[kpq]}_{\{i\}} \eea
with $C_2[G]$ standing for the second Casimir of the group $G$.

The second step thus corresponds to finding all the irreducible
representations of $SO(6)$ that contain a given representation of
$SO(5)$ under the decomposition $SO(6) \rightarrow SO(5)$.
Denoting $SO(5)$ irreps by their Dynkin labels, $[m,n]$,  standard
group theory analysis yields the KK towers
\be {\rm KK}_{[m,n]} =
\sum_{r=0}^{m}\sum_{s=0}^{n} \sum_{p=m-r}^\infty \left[r\pls
s,\,p,\,r\pls n\mis s\right] +\sum_{r=0}^{m-1} \sum_{s=0}^{n-1}
\sum_{p=m-r-1}^\infty \left[r\pls s\pls1 ,\,p,\,r\pls n\mis
s\right] \;. \label{KK56} \ee
 In particular for the smallest irreps one gets: $KK_{{\bf
1}=[0,0]} = \sum_{p=0}^\infty [0,p,0]$, $KK_{{\bf 5}=[1,0]} =
\sum_{p=1}^\infty [0,p,0] + \sum_{p=0}^\infty [1,p,1]$ , and $
KK_{{\bf 4}=[0,1]} = \sum_{n=0}^\infty [1,p,0] + [0,p,1]$. Notice
that any ambiguity in the lift, say, of the (pseudo-real) spinor
${\bf 4}$ of $SO(5)\approx Sp(4)$ to the complex ${\bf 4}$ of
$SO(6)\approx SU(4)$ or to its complex conjugate ${\bf 4}^*$ is
resolved by the infinite sum over KK recurrences. Once the $SO(6)$
R-symmetry quantum numbers are determined, the perturbative
spectrum of the superstring turns out to be encoded in \be {\cal
H}_{AdS} = {\cal H}_{sugra} + {\cal T}_{KK} {\cal T}_{susy}
\sum_\ell V^{L}_{\ell} \times V^{R}_{\ell} \label{adsspect}\ee
where ${\cal T}_{KK} = \sum_p [0,p,0]$ represents the KK tower
that, as indicated, boils down to a sum over scalar spherical
harmonics.
${\cal T}_{susy}$ represents the action of the 16 `raising'
supercharges $Q$ and $\bar{Q}$ with quantum numbers
$\{1/2;(1/2,0);[1,0,0]\}$ and $\{1/2;(0,1/2);[0,0,1]\}$,
respectively. $V^{L/R}_{\ell}$, defined in flat space, are to be
decomposed under $SO(4)\times SO(5)$. Formula (\ref{adsspect})
looks deceivingly simple, almost trivial, since the most
interesting and subtle information, the scaling dimension 
$\Delta_0$, at the HS enhancement point, is still missing.

Before addressing this crucial issue, two remarks are in order.
First we have tacitly assumed that there are no non-perturbative
states that can appear in the single-particle spectrum as a result
of strings or branes wrapping non trivial cycles
\cite{Witten:1998xy}. Indeed there are no such states with finite
mass, since the only non trivial cycles of $S^5$ are a 0-cycle (a
point) or a 5-cycle (the full space). Although there are no stable
type IIB 0-branes there are stable 5-branes of various kinds.
However they give rise to very massive objects (baryon vertices,
...) at small string coupling, \ie large $N$. Second, there can be
ambiguities in extrapolating the perturbative spectrum from large
radius, where KK technology is reliable but string excitations are
very massive, to small radius where HS symmetry is restored but
stringy geometry should replace more familiar concepts. We should
then appeal to the non-intersecting principle
\cite{Polyakov:2001af} that guarantees that any state identified
at large radius (strong 't Hooft coupling) can be smoothly
followed to weak coupling and viceversa. Indeed whenever the
dimensions of two (or more) operators with the same quantum
numbers start to approach one another level repulsion should
prevent them from actually coincide.

One can then start by identifying the string excitations that are
expected to become massless at the point of enhanced HS symmetry.
In particular the totally symmetric and traceless tensors of rank
$2\ell - 2$ at level $\ell>1$ appearing in the product of the
groundstates $V^{L}_{\ell} \times V^{R}_{\ell}$ become massless
and thus correspond to the sought for conserved currents on the
boundary if one assigns them $\Delta_0 = 2\ell$, that works
fine for $\ell=1$, too. The states with quantumm numbers
$\{2\ell;(\ell-1,\ell-1);[0,0,0]\}$ are HWS's of semishort
multiplets. $PSU(2,2|4)$ symmetry then fixes the scaling
dimensions of the other components. In practice, {\it one takes
$2^{16}$ birds with one stone!} Moreover the KK recurrences of
these states at floor $p$ arising from ${\cal T}_{KK}$ are
naturally assigned \be \Delta_0 = 2\ell + p\ee which represents
the $PSU(2,2|4)$ unitary bound for a spin $s= 2\ell - 2$ current
in the $SO(6)$ irrep with Dynkin labels $[0,p,0]$. It is
remarkable how simply assuming HS symmetry enhancement fixes the
AdS masses, \ie scaling dimensions, of a significant fraction of
the spectrum. Although experience with perturbative gauge theories
teaches us that even at this particularly symmetric point there
be operators / states well above the relevant $PSU(2,2|4)$
unitary bounds, the above very simple yet effective
mass formula turns out to be correct for all states with dimension
up to $\Delta_0 = 4$. Notice that `commensurability' of the
two contribution -- spin $s\approx \ell$ and KK `angular momentum'
$J \approx p$ -- suggests $R = \sqrt{\ap}$, for what this could
mean. In order to find a mass formula that could extend and
generalize the above one, it is convenient to take the BMN formula
as a hint. Although derived under the assumptions of large
$\lambda$ and $J$ there seems to be no serious problem in
extrapolating it to finite $J$ at vanishing $\lambda$, where
$\omega_n = 1$ for all $n$. Indeed (two-impurity) BMN operators
form $PSU(2,2|4)$ multiplets at finite $J$ and are thus amenable
to the extrapolation \cite{Beisert:2002bb}. The resulting formula
reads \be \Delta_0 = J + \nu \label{magic}\ee where $\nu = \sum_n
N_n$ is the number of oscillators applied to the `vacumm'
$|J=\mu\ap p^+\rangle$ and $J$ is a $U(1)$ R-charge yet to be
identified. The easiest way to proceed is to first `covariantize'
$SO(9)$ to $SO(10)$ and then decompose the latter into
$SO(8)\times U(1)_J$ where $SO(8)$ is the massless little group
and $U(1)_J$ is precisely the sought for R-charge. Although
cumbersome the procedure is straightforward and can be easily
implemented on a computer. Given the $SO(10)$ content of the flat
space string spectrum, equation~\eqref{magic}
uniquely determines the dimensions $\Delta_0$ of the superstring
excitations around $AdS_5\times S^5$ at the HS point. As an
illustration, let us consider the first few string levels:
\bea V_1 \eq [0,0,0,0,0]^1 \non \earel{\stackrel{SO(8)\times
SO(2)}{\rightarrow}} [0,0,0,0]^1_0 \non
\earel{\stackrel{\eqref{magic}}{\rightarrow}} [0,0,0,0]_1 \;, \eea
\bea V_2 \eq [1,0,0,0,0]^2-[0,0,0,0,0]^3 \non
\earel{\stackrel{SO(8)\times SO(2)}{\rightarrow}}
 [1,0,0,0]^2_0+
[0,0,0,0]^2_1+[0,0,0,0]^2_{-1} -[0,0,0,0]^3_0\label{second}\\
\earel{\stackrel{\eqref{magic}}{\rightarrow}}
 [1,0,0,0]_2+ [0,0,0,0]_1 \;,
\eea \bea  V_3 \eq [2,0,0,0,0]^3-[1,0,0,0,0]^4-[0,0,0,0,1]^{5/2}
\non \earel{\stackrel{SO(8)\times SO(2)}{\rightarrow}}
  [2,0,0,0]^3_0+
[1,0,0,0]^3_1+[1,0,0,0]^3_{-1} +[0,0,0,0]^3_0+ [0,0,0,0]^3_2 \nl
+[0,0,0,0]^3_{-2}
 -  [1,0,0,0]^4_0- [0,0,0,0]^4_1-[0,0,0,0]^4_{-1} \nl
-[0,0,0,1]^{5/2}_{1/2}-[0,0,1,0]^{5/2}_{-1/2} \non
\earel{\stackrel{\eqref{magic}}{\rightarrow}}
 [2,0,0,0]_3 +[1,0,0,0]_2
+[0,0,0,0]_1 -[0,0,1,0]_{3}-[0,0,0,1]_{2} \;. \eea

With the above assignments of $\Delta_0$, negative multiplicities are harmless 
since they cancel in the sum
over KK recurrences after decomposing $SO(10)$ w.r.t. $SO(4)\times SO(6)$. For 
these
low massive levels, the conformal dimensions determined
by~\eqref{magic} all saturate $SO(10)$ unitary bounds $\Delta_\pm=
1+ k + 2l + 3m + 2(p+q)\pm(p-q)/2$. At higher levels, starting
from a scalar singlet with $\Delta_0=3$ at level $\ell=5$, these bounds are
satisfied but no longer saturated. The correct conformal
dimensions are rather obtained from \eqref{magic}. Notice that the first 
fermionic primary appears at level
$\ell=3$ and has dimension $\Delta_0 = 11/2$.

Summarizing, the massive flat space string spectrum may be lifted
to $SO(10)\times SO(2)_\Delta$, such that breaking $SO(10)$ down
to $SO(8)\times SO(2)_J$ reproduces the original $SO(8)$ string
spectrum and its excitation numbers via the relation
\eqref{magic}. The results up to string level $\ell=5$ are
displayed in the following tables and organized under
$SO(10)\times SO(2)_\Delta$, with Dynkin labels
$[k,l,m,p,q]_{\Delta_0}$ and $[k,l,m,p,q]^* \equiv
[k,l,m,p,q]-[k\!-\!1,l,m,p,q]$.

     \noindent$\ell=1$ :
     {\footnotesize\begin{equation}\nonumber\begin{tabular}{l|l}
     $\Delta_0$ & ${\cal R}$ \\ \hline
     $1$ &   $
     [0, 0, 0, 0, 0]
     $\\ \hline
     \end{tabular}\end{equation}}

     \noindent$\ell=2$ :
     {\footnotesize\begin{equation}\nonumber\begin{tabular}{l|l}
     $\Delta_0$ & ${\cal R}$ \\ \hline
     $2$ &   $
     [1, 0, 0, 0, 0]^*
     $\\ \hline
     \end{tabular}\end{equation}}

     \noindent$\ell=3$ :
     {\footnotesize\begin{equation}\nonumber\begin{tabular}{l|l}
     $\Delta_0$ & ${\cal R}$ \\ \hline
     $3$ &   $
     [2, 0, 0, 0, 0]^*
     $\\ \hline
     $\frac{5}{2}$ &   $
     [0, 0, 0, 0, 1]
     $\\ \hline
     \end{tabular}\end{equation}}

     \noindent$\ell=4$ :
     {\footnotesize\begin{equation}\nonumber\begin{tabular}{l|l}
     $\Delta_0$ & ${\cal R}$ \\ \hline
     $4$ &   $
     [3, 0, 0, 0, 0]^*
     $\\ \hline
     $\frac{7}{2}$ &   $
     [1, 0, 0, 0, 1]^*
     $\\ \hline
     $3$ &   $
     [0, 1, 0, 0, 0]
     $\\ \hline
     \end{tabular}\end{equation}}

     \noindent$\ell=5$ :
     {\footnotesize\begin{equation}\nonumber\begin{tabular}{l|l}
     $\Delta_0$ & ${\cal R}$ \\ \hline
     $5$ &   $
     [4, 0, 0, 0, 0]^*
     $\\ \hline
     $\frac{9}{2}$ &   $
     [2, 0, 0, 0, 1]^*
     $\\ \hline
     $4$ &   $
     [0, 0, 1, 0, 0]
     +[1, 1, 0, 0, 0]^*
     $\\ \hline
     $\frac{7}{2}$ &   $
     [1, 0, 0, 0, 1]
     $\\ \hline
     $3$ &   $
     [0, 0, 0, 0, 0]
     $\\ \hline
     \end{tabular}\end{equation}}

\section{The spectrum of free ${\cal N}=4$ SYM }

In order to test the above prediction for the single-particle
superstring spectrum on $AdS_5\times S^5$ at the HS point with the
spectrum of free ${\cal N}=4$ SYM theory at large $N$, one has to
devise an efficient way of computing gauge-invariant single trace
operators. For $SU(N)$ gauge group this means taking care of the
ciclicity of the trace in order to avoid multiple counting 
\cite{Sundborg:1999ue, Sundborg:2000wp, Haggi-Mani:2000ru, Polyakov:2001af,
Bianchi:2003wx, Aharony:2003sx}.
Moreover one should discard operators which would vanish along the
solutions of the field equations and deal with the statistics of
the elementary fields properly. The mathematical tool one has to
resort to is Polya theorem \cite{Polya}
 that allows one to count `words' $A, B,
...$ of a given `length' $n$ composed of `letters' chosen from a given
alphabet $\{a_i\}$, modulo some symmetry operation: $A\approx B$
if $A=g B$ for $g\in {\cal G}$. Decomposing the discrete group
${\cal G}\subset {\cal S}_n$ into conjugacy classes whose
representatives $[g]=(1)^{b_1(g)} (1)^{b_1(g)}...(n)^{b_n(g)}$ are
characterized by the numbers $b_k(g)$ of cycles of length $k$,
Polya cycle index is given by \be {\cal P}_{\cal G}(\{a_i\}) =
{1\over |{\cal G}| }\sum_g \prod_{k=1}^n (\sum_i a_i^k)^{b_k(g)}
\ee For cyclic groups, ${\cal G}=Z_n$, conjugacy classes are labeled by
divisors $d$ of $n$, $[g]_d=(d)^{n/d}$, and the cycle index simply
reads \be {\cal P}_{Z_n}(\{a_i\}) = {1\over n }\sum_{d|n} {\cal
E}(d)(\sum_i a_i^d)^{n/d} \label{polyah}
\ee where ${\cal E}(d)$ is Euler totient
function which counts the number of elements in  the conjugacy
class $[g]_d$. ${\cal E}(d)$ equals the number of integers
relatively prime to and smaller than $d$, with the understanding
that ${\cal E}(1)=1$, and satisfies $\sum_{d|n} {\cal E}(d)=n$.

For ${\cal N}=4$ SYM the alphabet consists of the elementary
fields together with their derivatives $\{\partial^k \varphi,
\partial^k \lambda,
\partial^k F\}$, modulo the field equations, that transform in the {\it 
singleton} representation of
$PSU(2,2|4)$. As a first step, one computes the on-shell single
letter partition function ${\cal Z}_1(q) = Tr q^{\Delta_0}$, 
where $q$ keeps track of the `naive' scaling
dimension ${\Delta}_0$\footnote{Additional variables can introduced in order to
keep track of other quantum numbers and compute the character
valued partition function.}. For a free scalar of dimension
$\Delta_0 =1$ in $D=4$ \be {\cal Z}^{(s)}_1(q)= q{1-q^2\over
(1-q)^4} = q{1 + q \over (1-q)^3 } \ee where $-q^2$ removes the
(module of the) null descendant $\partial^2\varphi = 0$. For a
free Weyl fermion of dimension $\Delta_0 =3/2$
 \be
{\cal Z}^{(f)}_1(q)= 2 q^{3/2}{1-q\over (1-q)^4} = 2 q^{3/2}{1
\over (1-q)^3 } \ee where $-q$ removes the null descendant
$\spartial \lambda = 0$. For a free vector field or rather its
field strength of dimension $\Delta_0 =2$ \be {\cal Z}^{(v)}_1(q)= 2
q^2{(3 - 4q + q^2)\over (1-q)^4} = 2 q^2{3- q \over (1-q)^3 } \ee
where $-4q$ removes the  null descendants at level one
$\partial_\mu F^{\mu\nu} = \partial_\mu \tilde{F}^{\mu\nu} = 0$
and $+q^2$ takes care of the algebraic identities $\partial_\mu
\partial_\nu F^{\mu\nu} =
\partial_\mu \partial_\nu \tilde{F}^{\mu\nu} = 0$ at the next level.

Taking into account statistics, \ie computing the Witten
index $Tr(-)^F q^\Delta$, and setting $n_s = 6$, $n_f =
n_{\bar{f}} = 4$ and $n_v=1$ one gets \be {\cal Z}^{({\cal
N}=4)}_1(q)= 2 q{(3 +\sqrt{q})\over (1+\sqrt{q})^3} \ee for a
single (abelian) ${\cal N}=4$ vector multiplet, quite remarkably
${\cal Z}^{({\cal N}=4)}_1(q) = {\cal Z}^{(v)}_1(-\sqrt{q})$.

Plugging this in (\ref{polyah}) and removing the single-letter term ($n=1$),
as appropriate for an $SU(N)$ gauge group,  one finds
 \bea {\cal Z}^{({\cal N}=4)}(q) \eq
 \sum_{n=2}^\infty \sum_{n|d}
\frac{{\cal E}(d)}{n}\,\left[
\frac{2q(3+q^{\frac{d}{2}})}{(1+q^{\frac{d}{2}})^3}\right]^{\frac{n}{d}}
\label{polyahA}\\
\eq 21\,q^2 - 96\,q^{\frac{5}{2}} + 376\,q^3 -
1344\,q^{\frac{7}{2}} + 4605\,q^4 - 15456\,q^{\frac{9}{2}} +
  52152\,q^5
\nl - 177600\,q^{\frac{11}{2}} + 608365\,q^6 -
2095584\,q^{\frac{13}{2}} + 7262256\,q^7 -
25299744\,q^{\frac{15}{2}} \nl + 88521741\,q^8 -
310927104\,q^{\frac{17}{2}} + 1095923200\,q^9 -
3874803840\,q^{\frac{19}{2}} \nl + 13737944493\,q^{10} +
\mathcal{O}(q^{\frac{21}{2}}) \quad .\eea  At large $N$ mixing
with multi-trace operators is suppressed.

The next step is to identify super-primaries, which is
tantamount to passing ${\cal Z}^{({\cal N}=4)}(q)$ through an
Eratosthenes super-sieve, that removes super-descendants. This task
can be accomplished by first subtracting the contribution of 1/2 BPS
multiplets from (\ref{polyahA})
\bea {\cal Z}^{({\cal N}=4)}_{BPS}(q)\eq
 \sum_{n=2} ^\infty
\frac{q^n\bigbrk{n+2-(n-2)q^{\frac{1}{2}}}}{12(1+q^{\frac{1}{2}})^4}
\Bigl[ (n+1)(n+3)\bigbrk{(n+2) -3q^{\frac{1}{2}}(n-2)}+
\nl\hspace{4.5cm}{}
+q(n-1)(n-3)\bigbrk{3(n+2)-q^{\frac{1}{2}}(n-2)} \Bigr]
\nln \eq \frac{q^2\bigbrk{20+80q^{\frac{1}{2}}
+146q+144q^{\frac{3}{2}}+81q^2+24q^{\frac{5}{2}}+3q^3}}{(1-q)
(1+q^{\frac{1}{2}})^8} \;, \eea
from (\ref{polyahA}) and then dividing by \be {\cal T}_{SO(10,2)}(q) =
(1-q^2)\frac{(1-q^\frac{1}{2})^{16}}{
 (1-q)^{10}}.
 \ee
 that not only removes superconformal descendants generated by
 \be {\cal T}_{susy}(q) = \frac{(1-q^\frac{1}{2})^{16}}{
 (1-q)^{4}}.
 \ee
 but also the operators dual to the KK recurrences generated by
\be {\cal T}_{KK}(q) = \frac{(1-q^2)}{
 (1-q)^{6}}.
 \ee
 where the numerator implements the $SO(6)$ tracelessness condition.

One eventually finds
 \bea
{\cal Z}^{({\cal N}=4)}_{SO(10,2)}(q)\eq \left[
  {\cal Z}^{({\cal N}=4)}(q)-{\cal Z}^{({\cal N}=4)}_{BPS}(q)\right]/
{\cal T}_{SO(10,2)}(q) \label{numso10}\\
\eq
 q^2 + 100\,q^4 + 236\,q^5 - 1728\,q^{\frac{11}{2}} + 4943\,q^6
- 12928\,q^{\frac{13}{2}} \nl + 60428\,q^7 -
201792\,q^{\frac{15}{2}}
 + 707426\,q^8 - 2550208\,q^{\frac{17}{2}}
\nl + 9101288\,q^9 - 32568832\,q^{\frac{19}{2}} +
116831861\,q^{10} + \mathcal{O}(q^{\frac{21}{2}}) \;.\nn
\label{zso10A} \eea
 The expansion (\ref{numso10}) can be reorganized in the form
\bea {\cal Z}^{({\cal N}=4)}_{SO(10,2)}(q)\eq (q^1)^2+(10
q^2-q^3)^2
+(-16 q^{5/2} + 54 q^3 - 10 q^4)^2\nn\\
&& + (45 q^3 - 144 q^{\frac{7}{2}} + 210 q^4 + 16 q^{\frac{9}{2}}
- 54 q^5)^2+\ldots \;, \eea It is not difficult to recognize that
\be {\cal Z}^{({\cal N}=4)}_{SO(10,2)}(q) = \sum_{\ell}
V_{\ell}^L(q) \times V_{\ell}^R(q) \ee where $q$ keeps track of the
dimensions assigned via $\Delta_0 = J +\nu$ after lifting $SO(9)$ to
$SO(10)$. We thus find perfect agreement with the previously derived
string spectrum up to $\Delta_0=10$, and are confident that 
our assumptions and extrapolations to the
HS enhancement point are correct. The final result seems
to suggest holomorphic factorization of the string worldsheet
dynamics at this particularly symmetric point. The origin of the
$SO(10,2)$ spectrum symmetry calls for deeper understanding
possibly in connection with Bars's two-time formulation of the type
IIB superstring \cite{Bars:2002pe, Bars:2004dg}. 
In the following table we gather the spectrum of super-primaries of long 
multiplets in ${\cal N}=4$ SYM. For reasons of limited
space, we present the result only up to $\Delta_0={13}/{2}$. HWS of
long multiplets of $\superN=4$ are denoted by $
[j_L,j_R;k,p,q]_{L,B}^P$, where $(j_L,j_R)$ and $[k,p,q]$ indicate
the Dynkin labels of $SO(4)$ and $SO(6)$, respectively. In
addition we include parity $P$, described in
\cite{Beisert:2003jj}, and hypercharge (Intriligator's
`bonus' symmetry) $B$, the leading
order $U(1)_B$ charge in the decomposition $SU(2,2|4)=U(1)_B\times
PSU(2,2|4)$, and length $L$, the leading order number of letters /
partons. Furthermore, $P=\pm$ indicates a pair of states with
opposite parities while $+\mathrm{conj.}$ indicates a conjugate
state $[j_R,j_L;q,p,k]_{L,-B}^P$.

{\footnotesize\[\nonumber\begin{array}{l|l}
\Delta_0&\mathcal{R}\\
\hline
2&
[0,0;0,0,0]_{2,0}^{+}
\\\hline
3&
[0,0;0,1,0]_{3,0}^{-}
\\\hline
4&
2\mathord{\cdot}[4;0,0;0,0,0]_{4,0}^{+} +[0,0;1,0,1]_{4,0}^{-}
+2\mathord{\cdot}[4;0,0;0,2,0]_{4,0}^{+}
\\&\mathord{}
+([0,2;0,0,0]_{3,-1}^{-}+\mathrm{conj.})
+[1,1;0,1,0]_{3,0}^{\pm} +[2,2;0,0,0]_{2,0}^{+}
\\\hline
5&
4\mathord{\cdot}[0,0;0,1,0]_{5,0}^{-}
+2\mathord{\cdot}([0,0;0,0,2]_{5,0}^{+}+\mathrm{conj.})
+[0,0;1,1,1]_{5,0}^{\pm}
\\&\mathord{}
+2\mathord{\cdot}[5;0,0;0,3,0]_{5,0}^{-}
+2\mathord{\cdot}([0,2;0,1,0]_{4,-1}^{+}+\mathrm{conj.})
\\&\mathord{}
+([0,2;2,0,0]_{4,-1}^{-}+\mathrm{conj.})
+[1,1;0,0,0]_{4,0}^{\pm} +2\mathord{\cdot}[1,1;1,0,1]_{4,0}^{\pm}
\\&\mathord{}
+[1,1;0,2,0]_{4,0}^{\pm} +[2,2;0,1,0]_{3,0}^{-}
\\\hline
\sfrac{11}{2}&
2\mathord{\cdot}[0,1;1,0,0]_{5,-1/2}^{\pm}
+2\mathord{\cdot}[0,1;0,1,1]_{5,-1/2}^{\pm}
+[0,1;1,2,0]_{5,-1/2}^{\pm}
\\&\mathord{}
+2\mathord{\cdot}[1,2;0,0,1]_{4,-1/2}^{\pm}
+[1,2;1,1,0]_{4,-1/2}^{\pm}
+[2,3;1,0,0]_{3,-1/2}^{\pm}
+\mathrm{conjugates}
\\\hline
6&
2\mathord{\cdot}[0,0;0,0,0]_{4,0}^{+}
+2\mathord{\cdot}([0,0;0,0,0]_{5,-1}^{+}+\mathrm{conj.})
+5\mathord{\cdot}[0,0;0,0,0]_{6,0}^{+}
+3\mathord{\cdot}[0,0;1,0,1]_{6,0}^{+}
\\&\mathord{}
+6\mathord{\cdot}[0,0;1,0,1]_{6,0}^{-}
+9\mathord{\cdot}[0,0;0,2,0]_{6,0}^{+} +[0,0;0,2,0]_{6,0}^{-}
+3\mathord{\cdot}([0,0;0,1,2]_{6,0}^{-}+\mathrm{conj.})
\\&\mathord{}
+3\mathord{\cdot}[0,0;2,0,2]_{6,0}^{+} +[0,0;1,2,1]_{6,0}^{+}
+2\mathord{\cdot}[0,0;1,2,1]_{6,0}^{-}
+3\mathord{\cdot}[0,0;0,4,0]_{6,0}^{+}
\\&\mathord{}
+2\mathord{\cdot}([0,2;0,0,0]_{4,0}^{-}+\mathrm{conj.})
+3\mathord{\cdot}([0,2;0,0,0]_{5,-1}^{-}+\mathrm{conj.})
\\&\mathord{}
+4\mathord{\cdot}([0,2;1,0,1]_{5,-1}^{+}+\mathrm{conj.})
+2\mathord{\cdot}([0,2;1,0,1]_{5,-1}^{-}+\mathrm{conj.})
\\&\mathord{}
+4\mathord{\cdot}([0,2;0,2,0]_{5,-1}^{-}+\mathrm{conj.})
+([0,2;2,1,0]_{5,-1}^{+}+\mathrm{conj.})
+8\mathord{\cdot}[1,1;0,1,0]_{5,0}^{\pm}
\\&\mathord{}
+2\mathord{\cdot}([1,1;0,0,2]_{5,0}^{\pm}+\mathrm{conj.})
+4\mathord{\cdot}[1,1;1,1,1]_{5,0}^{\pm}
+2\mathord{\cdot}[1,1;0,3,0]_{5,0}^{\pm}
\\&\mathord{}
+([0,4;0,0,0]_{3,-1}^{+}+\mathrm{conj.})
+([0,4;0,0,0]_{4,-2}^{+}+\mathrm{conj.})
+2\mathord{\cdot}([1,3;0,1,0]_{4,-1}^{\pm}+\mathrm{conj.})
\\&\mathord{}
+5\mathord{\cdot}[2,2;0,0,0]_{4,0}^{+}
+2\mathord{\cdot}[2,2;0,0,0]_{4,0}^{-}
+2\mathord{\cdot}[2,2;1,0,1]_{4,0}^{-}
+4\mathord{\cdot}[2,2;0,2,0]_{4,0}^{+}
\\&\mathord{}
+[2,2;0,2,0]_{4,0}^{-}
+([2,4;0,0,0]_{3,-1}^{-}+\mathrm{conj.}) +[3,3;0,1,0]_{3,0}^{\pm}
+[4,4;0,0,0]_{2,0}^{+}
\\\hline
\frac{13}{2}&
4\mathord{\cdot}[0,1;0,0,1]_{5,+1/2}^{\pm}
+6\mathord{\cdot}[0,1;0,0,1]_{6,-1/2}^{\pm}
+12\mathord{\cdot}[0,1;1,1,0]_{6,-1/2}^{\pm}
\\&\mathord{}
+5\mathord{\cdot}[0,1;1,0,2]_{6,-1/2}^{\pm}
+[0,1;3,0,0]_{6,-1/2}^{\pm}
+5\mathord{\cdot}[0,1;0,2,1]_{6,-1/2}^{\pm}
\\&\mathord{}
+2\mathord{\cdot}[0,1;2,1,1]_{6,-1/2}^{\pm}
+[0,1;1,3,0]_{6,-1/2}^{\pm}
+[0,3;0,0,1]_{4,-1/2}^{\pm}
\\&\mathord{}
+[0,3;0,0,1]_{5,-3/2}^{\pm}
+2\mathord{\cdot}[0,3;1,1,0]_{5,-3/2}^{\pm}
+10\mathord{\cdot}[1,2;1,0,0]_{5,-1/2}^{\pm}
\\&\mathord{}
+8\mathord{\cdot}[1,2;0,1,1]_{5,-1/2}^{\pm}
+3\mathord{\cdot}[1,2;2,0,1]_{5,-1/2}^{\pm}
+2\mathord{\cdot}[1,2;1,2,0]_{5,-1/2}^{\pm}
\\&\mathord{}
+[1,4;1,0,0]_{4,-3/2}^{\pm}
+3\mathord{\cdot}[2,3;0,0,1]_{4,-1/2}^{\pm}
+2\mathord{\cdot}[2,3;1,1,0]_{4,-1/2}^{\pm}
+\mathrm{conjugates}\\
\hline
\end{array}
\]}

\section{HS symmetry and multiplets}

It is now time to decompose the spectrum of single-trace operators
in free ${\cal N}=4$ SYM at large $N$ or, equivalently, of type
IIB superstring on $AdS_5\times S^5$ extrapolated to the point of
HS symmetry, into HS multiplets in order to set the stage for
interactions and symmetry breaking. To this end we need to recall
some basic properties of the infinite dimensional HS
(super)algebra $\alhs(2,2|4)$, that extends the ${\cal N}=4$
superconformal algebra $\alPSU(2,2|4)$ ~\cite{Sezgin:2001zs, Sezgin:2001yf, 
Sezgin:2002rt, Brink:2000ag, Konstein:2000bi, Vasiliev:2001ur, 
Vasiliev:2004cm}. The latter can be realized
in terms of (super-)oscillators $\zeta_\Lambda = (y_a, \theta_A)$
with:
\[
 {}[y_a, \bar{y}^b] = \delta_a{}^b{} \;\quad , \qquad
\{\theta_A, \bar{\theta}^B\} = \delta^B{}_A \;\quad ,
\]
where $y_a, \bar{y}^b$ are bosonic oscillators with $a,b=1,...4$ a
Weyl spinor index of $\alSO(4,2)\sim \alSU(2,2)$ or, equivalently,
a Dirac spinor index of $\alSO(4,1)$, while
$\theta_A,\bar{\theta}^B$ are fermionic oscillators with
$A,B=1,...4$ a Weyl spinor index of $\alSO(6)\sim \alSU(4)$.

Generators of $\alPSU(2,2|4)$ are `traceless' bilinears
$\bar{\zeta}^\Sigma\zeta_\Lambda$ of superoscillators. In
particular, the `diagonal' combinations realize the compact
(R-symmetry) $\mathfrak{so}(6)$ and noncompact (conformal) 
$\mathfrak{so}(4,2)$ bosonic
subalgebras respectively, while the mixed combinations generate
supersymmetries:
\begin{eqnarray} &&J^a{}_b = \bar{y}^a y_b - \ft12 K \delta^a{}_b
\quad , \quad K = \ft12 \bar{y}^a y_a \quad , \quad \bar{\cal
Q}^a_A = \bar{y}^a \theta_A \quad ,
 \nn\\[1ex]
&&T^A{}_B = \bar{\theta}^A \theta_B - \ft12 B \delta^A_B \quad ,
\quad B = \ft12 \bar{\theta}^A \theta_A \quad , \quad {\cal Q}^A_a
= \bar{\theta}^A y_a \quad .
 \label{supconfgen}
\end{eqnarray}

The central element
\[ C \equiv  K+B = \ft12 \bar{\zeta}^\Lambda \zeta_\Lambda\;,
\] generates an abelian ideal that can be
modded out \eg by consistently assigning $C=0$ to the elementary
SYM fields and their (perturbative) composites. Finally, the
hypercharge $B$ is the generator of Intriligator's `bonus
symmetry{}'~\cite{Intriligator:1998ig, Intriligator:1999ff} dual
to the `anomalous' $U(1)_B$ chiral symmetry of type IIB
supergravity. It acts as an external
automorphism that rotates the supercharges.

The HS extension $\alhs(2,2|4)$ is roughly speaking generated by
odd powers of the above generators \ie combinations with equal odd
numbers of $\zeta_\Lambda$ and $\bar{\zeta}^\Lambda$. More
precisely, one first considers the enveloping algebra of
$\alPSU(2,2|4)$, which is an associative algebra and consists of
all powers of the generators, then restricts it to the odd part
which closes as a Lie algebra modulo the central charge $C$, and
finally quotients the ideal generated by $C$. It is easy to show
that $B$ is never generated in commutators (but $C$ is!) and thus
remains an external automorphism of $\alhs(2,2|4)$. Generators of
$\alhs(2,2|4)$ can be represented by `traceless' polynomials in
the superoscillators:
\begin{eqnarray}
\alhs(2,2|4) &=& \oplus_\ell\, {\cal A}_{2\ell+1} =
\sum_{\ell=0}^\infty \Big\{ {\cal J}_{2\ell+1}= P^{\Lambda_1\ldots
\Lambda_{2\ell+1}}_{\Sigma_1\ldots \Sigma_{2\ell+1}}\,
 \bar{\zeta}^{\Sigma_1}\!\dots\bar{\zeta}^{\Sigma_{2\ell+1}}\,
 \zeta_{\Lambda_1}\!\dots \zeta_{\Lambda_{2\ell+1}}
\Big\}\;,\label{phs}
\end{eqnarray}
with elements ${\cal J}_{2\ell+1}$ in ${\cal A}_{2\ell+1}$ at
level $\ell$ parametrized by traceless rank $(2\ell\!+\!1)$
(graded) symmetric tensors $P^{\Lambda_1\ldots
\Lambda_{2\ell+1}}_{\Sigma_1\ldots \Sigma_{2\ell+1}}$.
Alternatively, the HS algebra can be defined by identifying
generators differing by terms that involve $C$, i.e.~${\cal
J}\approx {\cal K}$ iff ${\cal J} - {\cal K}=\sum_ {k\ge 1} C^k
{\cal H}_k$.

To each element in $hs(2,2|4)$ with $\alSU(2)_L\times
\alSU(2)_R$ spins $(j_L,j_R)$ is associated an $\alhs(2,2|4)$ HS
currents and a dual HS gauge field in the AdS bulk with spins
$(j_L+\ft12,j_R+\ft12)$. The $\alPSU(2,2|4)$ quantum numbers of the
HS generators can be read off from (\ref{phs}) by expanding the
polynomials in powers of $\theta$'s up to 4, since $\theta^5=0$.
There is a single superconformal multiplet $\mult_{2\ell}$ at each
level $\ell\geq 2$. The lowest spin cases $\ell=0,1$, i.e.
$\hat{\mult}_{0,2}$, are special. They differ from the content of
doubleton multiplets $\mult_{0,2}$ by spin $s=0,1/2$ states
\cite{Sezgin:2001zs}. The content of (\ref{phs}) can then be
written as (tables 4,5 of \cite{Sezgin:2001zs})
\begin{eqnarray} \hat{\mult}_{0}&=& \big|\bar{{\bf
4}}_{[{\frac12},0]}+{\bf 1}_{[1,0]} \big|^2-
{\bf 1}_{[{\frac12},{\frac12}]}\nn\\
\hat{\mult}_{2}&=&  \big|{\bf 4}_{[{\frac12},0]}+{\bf 6}_{[1,0]}
+\bar{{\bf 4}}_{[{\frac32},0]}+{\bf 1}_{[2,0]} \big|^2\nn\\
 \mult_{2\ell}&=&
 \big| {\bf 1}_{[\ell-1,0]}+ {\bf 4}_{[\ell-{\frac12},0]}+
   {\bf 6}_{[\ell,0]}+ \bar{{\bf 4}}_{[\ell+{\frac12},0]}+
   {\bf 1}_{[\ell+1,0]} \big|^2 \;,   \qquad \ell\geq2\;,
\end{eqnarray} with ${\bf r}_{[j_L+\ft12,j_R+\ft12]}$ denoting the
$\alSU(4)$ representation~${\bf r}$ and the $\alSU(2)^2$ spins of the
 HWS's. Complex conjugates are given by
conjugating $\alSU(4)$ representations and exchanging the spins
$j_L\leftrightarrow j_R$. The product is understood in $\alSU(4)$
while spins simply add.
 The highest spin state ${\bf 1}_{[\ell+1,\ell+1]}$ corresponds to the state
 $y^{2\ell+1}\bar{y}^{2\ell+1}$ with no $\theta$'s, 
${\bf 4}_{[\ell+{\frac12},\ell+1]}$,
 $\bar{{\bf 4}}_{[\ell+1,\ell+{\frac12}]}$ to 
$y^{2\ell}\bar{y}^{2\ell+1} \theta^A$,
$y^{2\ell+1}\bar{y}^{2\ell} \bar{\theta}_A$, and so on. For
$\ell=0,1$, states with negative $j_L,j_R$ should be deleted. In
addition we subtract the current ${\bf 1}_{[{\frac12},{\frac12}]}$
at $\ell=0$ associated to $C$. In the ${\cal N}=4$ notation
introduced in Appendix~A,  ${\mult}_{2\ell}$ corresponds to the
semishort multiplet ${\mult}_{[000][\ell-1^*,\ell-1^*]}^{2\ell,0}$.

The singleton representation $\mult_{[0,1,0][0,0]}^{1,0}$ of
$\alPSU(2,2|4)$ truns out to be the fundamental representation of
$\alhs(2,2|4)$, too. Its HWS $|Z\rangle$, or simply $Z$ \ie the
ground-state or `vacuum', which is not to be confused with the
trivial $\alPSU(2,2|4)$ invariant vacuum $|0\rangle$, is one of
the complex scalars, say, $Z=\varphi^5 + i \varphi^6$. Showing
that the singleton is an irreducible representation of
$\alPSU(2,2|4)$ is tantamount to showing that any state $A$ in
this representation can be found by acting on the vacuum $Z$, or
any other state $B$, with a sequence of superconformal generators
(\ref{supconfgen}). Looking at the singleton as an irrep of
$\alhs(2,2|4)$ one sees an important difference: the sequence of
superconformal generators \footnote{Without loss of generality we
may assume the length of the sequence to be odd; for an even
sequence we may append an element of the Cartan subalgebra,
e.g.~the dilatation generator.} is replaced by a single HS
generator $\mathcal{J}_{A\bar B}$. Therefore any $A$ in the
$\alhs(2,2|4)$ singleton multiplet can be reached in a single step
 from any other one $B$ as can be shown
by noticing that, since the central charge $C$ commutes with all
generators and annihilates the vacuum $Z$, a non-trivial sequence
in $({\cal A}_1)^{2\ell+1}$ belongs to ${\cal A}_{2\ell+1}$. This
property will be crucial in proving the irreducibility of
YT-pletons with respect to the HS algebra. Let us then consider
the tensor product of $L$ singletons, \ie $L$ sites or partons.
The generators of $\alhs(2,2|4)$ are realized as diagonal
combinations: \[ {\cal J}_{2\ell+1}\equiv \sum_{s=1}^L\, {\cal
J}^{(s)}_{2\ell+1} \label{genl} \] with ${\cal J}_{2\ell+1}^{(s)}$
HS generators acting on the $s^{\rm th}$ site. The tensor product
of $L\geq 1$ singletons  is generically reducible not only under
$\alPSU(2,2|4)$ but also under $\alhs(2,2|4)$. This can be seen by
noticing that the HS generators (\ref{genl}), being completely
symmetric, commute with (anti)symmetrizations of the indices. In
particular, the tensor product decomposes into a sum of
representations characterized by Young tableaux $YT$ with $L$
boxes.

To prove irreducibility of $L$-pletons associated to a specific
YT under $\alhs(2,2|4)$, it is enough to show that any state in
the $L$-pleton under consideration can be found by acting on the
relevant HWS with HS generators. Let us start by considering
states belonging to the totally symmetric tableau. The simplest
examples of such states are those with only one `impurity'
i.e.~$AZ\ldots Z+\mbox{symm.}$. Using the fact that any SYM letter
$A$ can be reached from the HWS $Z$ by means of a single
$\alhs(2,2|4)$ generator $\mathcal{J}_{A\bar Z}$ we have $({\cal
J}_{A\bar Z}Z)Z\ldots Z+{\rm symm.}\approx{\cal
J}_{A\bar{Z}}(Z^L)$ that shows the state is a HS descendant.
 The next simplest class is given by states with
two impurities $ABZ\ldots Z+{\rm symm.}$. Once again these states
can be written as ${\cal J}_{A\bar Z}{\cal J}_{B\bar Z}(Z^L)$, up
to the one impurity descendants $({\cal J}_{A\bar Z}{\cal
J}_{B\bar Z}Z)Z\ldots Z$ of the type already found. Proceeding in
this way one can show that all states in the completely symmetric
tensor of $L$ singletons can be written as HS descendants of the
vacuum $Z^L$.

The same arguments hold for generic tableaux. For example, besides
the descendants ${\cal J}_{A\bar{Z}} (Z^L)$ of $Z^L$ there are
$L-1$ ``one impurity{}'' multiplets of states associated to the
$L-1$ Young tableaux with $L-1$ boxes in the first row and a
single box in the second one\footnote{As we will momentarily see,
HS multiplets of this kind are absent for ${\cal N}=4$ SYM
theories with semisimple gauge group. At any rate, they are
instrumental to illustrate our point.}. The vacuum state of HS
multiplets associated to such tableaux can be taken to be
$Y_{(k)}\equiv Z^{k} Y Z^{L-k-1}-Y Z^{L-1}$ with $k=1,\ldots,L-1$,
where $Y=\varphi^3 + i \varphi^4$ is another (complex) scalar. Any
state with one impurity $Z^{k} A Z^{L-k-1}-A Z^{L-1}$ with
$k=1,\ldots,L-1$ can be found by acting on $Y_{(k)}$ with the HS
generator ${\cal J}_{A\bar Y}$, where ${\cal J}_{A\bar Y}$ is the
HS generator that transforms $Y$ into $A$ (and annihilates $Z$).
Notice that the arguments rely heavily on the fact that any two
states in the singleton are related by a one-step action of a HS
generator. This is not the case for the ${\cal N}=4$ SCA, and
indeed the completely symmetric tensor product of $L$ singletons
is highly reducible with respect to $\alPSU(2,2|4)$.

The on-shell field content of the singleton representation of
$\alPSU(2,2|4)$ is encoded in the single-letter partition function
${\cal Z}_1(q)= \Yboxdim4pt {\cal Z}_{\yng(1)}(q)$. As previously described,
the spectrum
of single-trace operators in ${\cal N}=4$ SYM theory with $SU(N)$
gauge group at large $N$ can be extracted from the generating
function\cite{Sundborg:1999ue, Sundborg:2000wp, Haggi-Mani:2000ru, 
Polyakov:2001af, Bianchi:2003wx, Aharony:2003sx }
\[
\Yboxdim4pt {\cal Z}(q)= \sum_{n\geq 2}\, {\cal Z}_n (q)=
\sum_{n\geq 2,d|n}\, \frac{{\cal E}(d)}{n}\,{\cal
Z}_{\yng(1)}\,(q^d)^{\frac{n}{d}} \;, \label{polyah2}
\]
of cyclic words of length $L=n$. Observe that $\Yboxdim4pt 
{\cal Z}_{\yng(1)}\,(q^d)$ can be rewritten
as the alternating sum over length-$d$ Young tableaux of hook
type:
\[
\Yboxdim4pt {\cal Z}_{\yng(1)}\,(q^d) ~=~ {\cal
Z}_{\yng(5)\cdot\cdot\yng(2)}(q)~-~ {\cal
Z}_{\yng(4,1)^{\cdot\cdot\yng(2)}}(q)~+~ {\cal
Z}_{\yng(3,1,1)^{^{\cdot\cdot\yng(2)}}}(q)~-~ {\cal
Z}_{\yng(2,1,1,1)^{^{^{\cdot\cdot\yng(2)}}}}(q) ~+~ \ldots \;.
\label{hook}
\]
 Plugging this expansion into
(\ref{polyah2}), we find for the first few cases:
\begin{eqnarray}
\Yboxdim4pt {\cal Z}_2 &=& \Yboxdim4pt
{\cal Z}_{\yng(2)}\;,\nn\\[1ex]
{\cal Z}_3 &=& \Yboxdim4pt {\cal Z}_{\yng(3)}+{\cal
Z}_{\yng(1,1,1)}
\;,\nn\\[1ex]
{\cal Z}_4 &=& \Yboxdim4pt {\cal Z}_{\yng(4)}+ {\cal
Z}_{\yng(2,1,1)}+{\cal Z}_{\yng(2,2)}
\;,\nn\\[1ex]
{\cal Z}_5 &=& \Yboxdim4pt {\cal Z}_{\yng(5)}+{\cal
Z}_{\yng(3,2)}+ 2\, {\cal Z}_{\yng(3,1,1)}+ {\cal
Z}_{\yng(2,2,1)}+ {\cal Z}_{\yng(1,1,1,1,1)}\;, \qquad\mbox{etc.}
\label{hsD}
\end{eqnarray}

As anticipated, only a subset of YT's, those compatible with cyclicity
of the trace, enters in (\ref{hsD}). In particular, HS multiplets
associated to the tableaux $\Yboxdim6pt\yng(1,1)$,
$\Yboxdim6pt\yng(2,1)$, two out of the three of type
$\Yboxdim6pt\yng(2,1,1)$, and so on, are projected out. The
content of the various components in (\ref{hsD}) can be derived
from  
\be
\Yboxdim4pt {\cal Z}_{\yng(2)} = \Yboxdim4pt \frac{1}{
2!}\left[{\cal Z}_{\yng(1)}\,(t)^2+ {\cal
Z}_{\yng(1)}\,(t^2)\right]
\ee
\be
\Yboxdim4pt {\cal Z}_{\yng(3)} = \Yboxdim4pt \frac{1}{
3!}\left[{\cal Z}_{\yng(1)}\,(t)^3+ 3\,{\cal
Z}_{\yng(1)}\,(t^2){\cal Z}_{\yng(1)}\,(t)
+2\,{\cal Z}_{\yng(1)}\,(t^3)\right]
\ee
\be
\Yboxdim4pt {\cal Z}_{\yng(1,1,1)} = \Yboxdim4pt \frac{1}{
3!}\left[{\cal Z}_{\yng(1)}\,(t)^3- 3\,{\cal
Z}_{\yng(1)}\,(t^2){\cal Z}_{\yng(1)}\,(t)+2\,{\cal
Z}_{\yng(1)}\,(t^3)\right]
\ee
\be
\Yboxdim4pt {\cal Z}_{\yng(4)} = \Yboxdim4pt \frac{1}{
4!}\left[{\cal Z}_{\yng(1)}\,(t)^4+ 6\,{\cal
Z}_{\yng(1)}\,(t^2){\cal Z}_{\yng(1)}\,(t)^2+3\,{\cal
Z}_{\yng(1)}\,(t^2)^2+ 8\,{\cal Z}_{\yng(1)}\,(t^3){\cal
Z}_{\yng(1)}\,(t)+6\,{\cal
Z}_{\yng(1)}\,(t^4)\right]
\ee
\be
\Yboxdim4pt {\cal Z}_{\yng(2,2)} = \Yboxdim4pt \frac{1}{
4!}\left[2\, {\cal Z}_{\yng(1)}\,(t)^4+ 6\,{\cal
Z}_{\yng(1)}\,(t^2)^2-8\,{\cal Z}_{\yng(1)}\,(t^3)
\,{\cal Z}_{\yng(1)}(t)\right]
\ee
\be
\Yboxdim4pt {\cal Z}_{\yng(2,1,1)}= \Yboxdim4pt
\frac{1}{4!}\left[3\, {\cal Z}_{\yng(1)}\,(t)^4-6\,{\cal
Z}_{\yng(1)}\,(t^2) {\cal Z}_{\yng(1)}\,(t)^2- 3\,{\cal
Z}_{\yng(1)}\,(t^2)^2+6\,{\cal Z}_{\yng(1)}\,(t^4)\right]
\label{ds} \;.
\ee
that can be explicitly verified with the use of (\ref{hook}).

Under the superconformal group $\alPSU(2,2|4)$, the HS multiplet
${\cal Z}_{YT}$, associated to a given Young tableau $YT$ with $L$
boxes, decomposes into an infinite sum of multiplets. The HWS's
can be found by computing ${\cal Z}_{YT}$ and eliminating the
superconformal descendants by passing ${\cal Z}_{YT}$ through a
sort of Eratosthenes (super) sieve~\cite{Bianchi:2003wx}. In the
$\alPSU(2,2|4)$ notation $\mult^{\Delta,B}_{[j,\jb][q_1,p,q_2]}$
of the Appendix one finds for $L=2,3$

\< \Yboxdim4pt {\cal Z}_{\yng(2)}\eq
\sum_{n=0}^\infty\mult^{2n,0}_{[-1+n^\ast,-1+n^\ast][0,0,0]} \;,
\nln \Yboxdim4pt {\cal Z}_{\yng(3)} \eq \sum_{n=0}^\infty
c_n\left[
\mult^{1+n,0}_{[-1+\frac{1}{2}n^\ast,-1+\frac{1}{2}n^\ast][0,1,0]}
+ \bigbrk{\mult^{\frac{11}{2}+n,
\frac{1}{2}}_{[\frac{3}{2}+\frac{1}{2}n^\ast,1+\frac{1}{2}n^\ast][0,0,1]}
+\mbox{h.c.}}\right] \nl +\sum_{m=0}^\infty\sum_{n=0}^\infty
c_n\left[
\mult^{4+4m+n,1}_{[1+2m+\frac{1}{2}n^\ast,\frac{1}{2}n][0,0,0]}
+\mult^{9+4m+n,1}_{[\frac{7}{2}+2m+\frac{1}{2}n^\ast,
\frac{3}{2}+\frac{1}{2}n][0,0,0]} +\mbox{h.c.}\right] \;, \nln
\Yboxdim4pt {\cal Z}_{\yng(1,1,1)} \eq \sum_{n=0}^\infty c_n\left[
\mult^{4+n,0}_{[\frac{1}{2}+\frac{1}{2}n^\ast,
\frac{1}{2}+\frac{1}{2}n^\ast][0,1,0]} +\bigbrk{ \mult^{
\frac{5}{2}+n, \frac{1}{2}}_{[\frac{1}{2}n^\ast,-
\frac{1}{2}+\frac{1}{2}n^\ast][0,0,1]} +\mbox{h.c}}\right] \nl
+\sum_{m=0}^\infty\sum_{n=0}^\infty c_n\left[
\mult^{6+4m+n,1}_{[2+2m+\frac{1}{2}n^\ast,\frac{1}{2}n][0,0,0]} +
\mult^{7+4m+n,1}_{[\frac{5}{2}+2m+\frac{1}{2}n^\ast,
\frac{3}{2}+\frac{1}{2}n][0,0,0]} +\mbox{h.c.}\right]\;.
\label{456} \>

The multiplicities $c_n\equiv 1+[n/6]-\delta_{n,1~{\rm mod}~6}$
with $[m]$ the integral part of $m$, of $\alPSU(2,2|4)$ multiplets
inside $\alhs(2,2|4)$ count the number of ways one can distribute
HS descendants among the boxes in the tableaux.

In addition to the $\ft12$-BPS multiplet with $n=0$, the symmetric doubleton
$\Yboxdim4pt{\cal Z}_{\yng(2)}$ , corresponding to the quadratic
Casimir $\delta_{ab}$, contains the semishort multiplets of conserved HS
currents $\mult_{2n}$. The antisymmetric doubleton
$\Yboxdim4pt{\cal Z}_{\yng(1,1)}$  is ruled out by cyclicity of
the trace, cf.~\eqref{hsD}.  The `symmetric tripleton'
$\Yboxdim4pt{\cal Z}_{\yng(3)}$, corresponding to the cubic
Casimir $d_{abc}$, contains the first KK recurrences of twist 2
semishort multiplets, the semishort-semishort series $\mult_{\pm
1,n}$ starting with fermionic primaries and long-semishort
multiplets. The antisymmetric tripleton $\Yboxdim4pt{\cal
Z}_{\yng(1,1,1)}$, corresponding to the structure constants
$f_{abc}$, on the other hand contains the Goldstone multiplets
that merge with twist 2 multiplets to form long multiplets when
the HS symmetry is broken. In addition, fermionic
semishort-semishort multiplets and long-semishort multiplets also
appear.

\section{Glimpses of ${\mathcal La \: Grande \: Bouffe}$}

The problem of formulating the dynamics of HS fields dates back to
Dirac, Wigner, Fierz and Pauli. In the massless bosonic case,
Fronsdal has been able to write down linearized field equations
for totally symmetric tensors $\varphi^{(\mu_1...\mu_s)}$ that in
$D=4$ arise from the quadratic action \cite{Fronsdal:1978rb,
Vasiliev:2001ur, Vasiliev:2004qz}
\begin{eqnarray}
S_2^{(s)} &&= {1\over 2} (-)^s\int d^4x \, \{\partial_\nu
\varphi_{\mu_1...\mu_s}\partial^\nu \varphi^{\mu_1...\mu_s}
\\
&&- {s(s-1)\over 2} \partial_\nu
\varphi^{\lambda}{}_{\lambda\mu_3...\mu_s}\partial^\nu
\varphi_\rho^{\rho\mu_3...\mu_s} + s(s-1)\partial_\nu
\varphi^{\lambda}{}_{\lambda\mu_3...\mu_s}\partial_\rho \varphi^
{\nu\rho\mu_3...\mu_s}
\nonumber\\
&&- s \partial_\nu \varphi^{\nu}{}_{\mu_2...\mu_s}\partial_\rho
\varphi^{\rho\mu_2...\mu_s} - {s(s-1)(s-2)\over 4}\partial_\nu
\varphi^{\nu\rho}{}_{\rho\mu_2...\mu_s}\partial_\lambda
\varphi_{\sigma}{}^{\lambda\sigma\mu_2...\mu_s} \}\nonumber
\end{eqnarray}
upon imposing `doubly tracelessness' $\eta^{\mu_1\mu_2}
\eta^{\mu_3\mu_4}\varphi_{\mu_1...\mu_s}=0$. HS gauge invariance
correspond to transformations
$$
\delta \varphi_{\mu_1...\mu_s} = \partial_{(\mu_1}
\epsilon_{\mu_2...\mu_s)} \label{hsaction}
$$
with traceless paramaters
$\eta^{\mu_1\mu_2}\epsilon_{\mu_1...\mu_{s-1}}=0$. Fang and Frosdal have then 
extended the analysis to fermions \cite{Fang:1978wz}, while  Singh and Hagen
formulated similar equations for massive fields with the help of
auxiliary fields, that reduce to Frondal's or Fang-Fronsdal's in
the massless limit upon removing certain auxiliary
fields\cite{Singh:1974qz, Singh:1974rc}. String theory in flat
spacetime can be considered as a theory of an infinite number of
HS gauge fields of various rank and (mixed) symmetry in a broken
phase. At high energies these symmetries should be restored
resulting in a new largely unexplored phase. 

Upon coupling to (external) gravity, the presence of the Weyl
tensor in the variation of the action for $s>2$, resulting from
the Riemann tensor in the commutator of two covariant derivatives,
spoils HS gauge invariance even at the linearized level except for
spin $s\le 2$, where at most the Ricci tensor appears. Problems
with interactions for HS gauge fields in flat spacetime are to be
expected since the Coleman - Mandula theorem and its generalization
by Haag - Lopusanski - Sohnius
lead to a trivial S-matrix whenever the (super)Poincar\`e group is
extended by additional spacetime generators such as HS symmetry
currents. Moreover closure of the HS algebra requires an infinite
tower of symmetries as soon as HS fields with $s>2$ enter the
game. A completely new approach to the interactions, if any, is to
be expected in order to deal with an infinite number of HS fields
and arbitrarily high derivatives.

According to Fradkin and Vasiliev, the situation improves
significantly when the starting point is taken to be a maximally
symmetric AdS space\footnote{Results for dS space formally obtain
by analytic continuation.} with non-vanishing cosmological
constant $\Lambda=-(D-2)(D-1)/R^2$ rather than flat spacetime. One
can then use the HS analogue of the MacDowell, Mansouri, Stelle,
West (MDMSW) $SO(d,2)$ formulation of gravity in order to keep HS
gauge symmetry manifest and compactly organize the resulting
higher derivative interactions and the associated non-locality.
Misha Vasiliev \cite{Vasiliev:2001ur, Vasiliev:2001wa, Alkalaev:2002rq,
Vasiliev:2003ev, Vasiliev:2004qz, Vasiliev:2004cm} has been able to pursue 
this program till the very end,
\ie at the fully non-linear level, for massless bosons in $D=4$.

The AdS/CFT correspondence at the HS enhancement point seems
exactly what the  doctor ordered. At generic radius $R$,
superstring theory describes HS fields in a broken phase. At some
critical radius, Vasiliev's equations govern the dynamics of the exactly 
massless phase. Here
$\Lambda$ plays a double role. On the one hand it suppresses higher
derivative interactions, very much like the string scale $M_s =
1/\sqrt{\ap}$ does in string theory. On the other hand it allows
one to define a generalized $SO(d,2)$ curvature (the bulk is
$D=d+1$ dimensional) that vanishes exactly for AdS.


In MDMSW formulation, one treats gravity with cosmological
constant in $D=d+1$ dimensions as an $SO(d,2)$ gauge theory with a
`compensator' $V^A(x)$ ($A=0,1,...d-1,d,d+1$) such that $\eta_{AB}
V^A V^B = - R^2$ with $\eta_{AB}=(-,+,...,+|+,-)$. To this end,
one extends the familiar frame and connection one-forms
$$
 e^a(x) = dx^m e^a_m(x) \quad , \quad
\omega^{ab}(x) = dx^m \omega^{ab}_m(x)
$$
with $m,n=0,1,...d$ to
$$
E^A = dV^A + \omega^A{}_B V^B = D V^A \ ,
$$
so that $E_A V^A = 0$, and
$$
\omega^{AB} = \omega^{AB}_L - \Lambda(E^A V^B - E^B V^A)
$$
where $\omega^{AB}_L$ is the generalized Lorentz $SO(d,1)$
connection, such that $D_L V^A = dV^A + \omega_L^A{}_B V^B = 0$.
In the ` unitary gauge', $V^A=R \delta^A_{d+1}$, $\omega^{ab}_L =
\omega^{ab}$ and $e^a = \omega^{aA} V_A$. The generalized
curvature two-forms
$$
R^A{}_B = d\omega^A{}_B + \omega^A{}_C \wedge \omega^C{}_B
$$
contain a longitudinal `torsion' part $R^A=R^A{}_BV^B = D E^A$ and
a transverse `Lorentz' $SO(d,1)$ part. For (A)dS
$$
R^A{}_B = 0
$$
with $rk(E^A_m) = d+1$ in order for the existence of a
non-degenerate metric tensor $g_{mn} = \eta_{AB} E^A_m E^B_n$. The
MDMSW action in $D=d+1$ dimensions is given by
$$
S = - {1\over 4 \kappa \sqrt{\Lambda}} \int_{M_{d+1}}
\epsilon_{A_0A_1...A_{d+1}} R^{A_0 A_1}\wedge R^{A_2 A_3}\wedge
E^{A_4}\wedge ...\wedge E^{A_d} V^{A_{d+1}}
$$
$S$ is manifestly invariant under diffeomorphisms and, thanks to
the compensator $V^A$, under $SO(d,2)$ gauge transformations
$$
\delta \omega^{AB} = D\varepsilon^{AB} \quad , \quad \delta V^A =
- \varepsilon^{AB} V_B
$$
Fixing the $SO(d,2)$ gauge requires a compensating diffeomorphism
so that
$$
\delta^\prime V^A = 0 = \delta_\varepsilon V^A + \delta_\xi V^A =
\xi^m E^A_m - \varepsilon^{AB} V_B
$$
The most symmetric solutions of the field equations are 'flat
connections' $\omega^{AB}_0$ and correspond to AdS. Global
symmetries very much like Killing vectors satisfy $D_0
\varepsilon^{AB} =0$, integrability follows from the
zero-curvature condition $R_0^A{}_B = 0$.

Generalization to massless HS gauge fields \cite{Vasiliev:2004qz}
is conveniently
described by a set of one-forms $\omega^{a_1...a_{s-1},
b_1...b_t}$ \`a la De Wit and Friedman \cite{deWit:1979pe} 
with $t\le s-1$ and $a_i,
b_i=0,...,d$, that represent two-row Young tableaux of $SO(d,1)$
with $s-1$ and $t$ boxes respectively. Dynamical fields are frame
type $\varphi^{a_1...a_{s}}= e^{n(a_s}\omega_n^{a_1...a_{s-1})}$
totally symmetric and (doubly) traceless tensors. Higher $t$
generalized connections are auxiliary and expressible in terms of
order $t$ derivatives of $\varphi^{a_1,...a_{s}}$. In the
$SO(d,2)$ invariant formulation, the generalized connections are
$\omega^{A_1,...A_{s-1}, B_1,...,B_t}$. One can define the
linearized HS curvature two-forms
$$
R_1^{A_1...A_{s-1}, B_1...B_{s-1}} = D_0\omega^{A_1...A_{s-1},
B_1...B_{s-1}}
$$
where $D_0$ is the background $SO(d,2)$ covariant derivative with
flat connection $\omega_0^{AB}$. The $SO(d,2)$ covariant form of
the (quadratic) action for spin $s$ is
\begin{eqnarray} S_2^{(s)} =
{1\over 2} \sum_{p=0}^{s-2} a(s,p) \epsilon_{A_0A_1...A_{d+1}}
\int_{M_{d+1}} E^{A_4}\wedge ...\wedge E^{A_d}
V^{A_{d+1}}V_{C_1}...V_{C_{2(s-2-p)}}\wedge
\\
R^{A_0}{}_{B_1...B_{s-2}}{}^{,A_1C_1...C_{s-2-p}}{}_{D_1...D_p}\wedge
R^{A_2B_1...B_{s-2},A_3C_{s-p-1}...C_{2(s-2-p)}D_1...D_p}
\end{eqnarray}
where $ a(s,p) = b(s) (-\Lambda)^{-(s-p-1)} {[d-5 + 2(s-p-2)]!!
(s-p-1)/(s-p-2)!}$ and $b(s)$ is fixed by the so-called `extra
fields decoupling condition', that prevents the propagation of the
auxiliary lower spin fields with $t<s$  \cite{Vasiliev:2004qz}. Linearized HS
gauge invariance under
$$
\delta \omega^{A_1...A_{s-1}, B_1...B_{s-1}}=
D_0\varepsilon^{A_1...A_{s-1}, B_1...B_{s-1}}
$$
is a consequence of the zero-curvature condition $R_0^A{}_B = 0$.
In higher dimensions \eg $D=5$ mixed symmetry tensors and spinors
may appear and in fact do so as a consequence e.g. of
supersymmetry.


Minimal (bosonic) HS symmetry can be defined as the symmetry of a
massless scalar field theory living on the $d(=D-1)$ dimensional
boundary of AdS \cite{Vasiliev:2004qz}. In the holographic perspective, global 
symmetries
on the boundary correspond to (global remnants of) local
symmetries in the bulk. HS symmetry should correspond to the
global symmetry of the maximally symmetric background.  In the AdS
vacuum, HS symmetry is generically broken to the (super)conformal
symmetry, except possibly for the point of enhanced HS symmetry,
that should correspond to some small curvature radius.

Bosonic HS symmetry algebras admit generators
$T_{A_1...A_s,B_1...B_s}$ in the two-row Young tableaux of
$SO(d,2)$ with $s=t$. Generalized commutation relations with the
$SO(d,2)\subset HS(d,2)$ generators $T_{A,B}$ take the obvious
form
$$
{} [T^C{}_D,T_{A_1...A_n,B_1...B_n}] = \delta^C{}_{A_1}
T_{D...A_n,B_1...B_n} + ...
$$
For later purposes it is convenient to introduce two sets of
oscillators $Y^A_i$ with $i=1,2$ that satisfy
$$
{} [Y^A_i, Y^B_j]_* = \eta^{AB} \epsilon_{ij} 
$$
where the Moyal-Weyl $*$-product is defined by
$$
(f*g)(Y) = \int {dT dS \over\pi^{2(d+2)}} f(Y+S) g(Y+T)
\exp(-2S\cdot T)$$
 The associative Weyl algebra ${\cal A}_{d+2}$
of polynomials
$$
P_n(Y) = \sum_{\{A_p,B_q\}} \varphi_{A_1..A_m,B_1...B_n}
Y_1^{A_1}...Y_1^{A_m}Y_2^{B_1}...Y_2^{B_n}
$$
contains $so(n,m)\oplus sp(2)$. Moreover, although the subalgebra
${\cal A}_{d+2}$ of $sp(2)$ singlets is not simple one can mod out
the ideal ${\cal I}$ generated by elements of the form
$a^{ij}*t_{ij}= t_{ij}*a^{ij}$. Only traceless two-row Yang
tableaux appear in the expansion. The Lie algebra with the
commutator in ${\cal S}/{\cal I}$ has a real form denoted by
$hu(1/sp(2)[n,m])$, which is the algebra of HS symmetry. In $D=4$
HS gauge fields can be compactly assembled into a connection
`master field' or simply `master connection'
$$
\omega(Y|x) = \sum_{n=0}^\infty \varphi_{A_1..A_n,B_1...B_n}
Y_1^{A_1}...Y_1^{A_n}Y_2^{B_1}...Y_2^{B_n}
$$
Gauge transformations are compactly encoded in
$$
\delta\omega(Y|x)= D\varepsilon(Y|x)\equiv d\varepsilon(Y|x) +
[\omega(Y|x),\varepsilon(Y|x)]_*
$$
The curvature `master field' is defined as usual
$$
R(Y|x) = d\omega(Y|x) + \omega(Y|x)\wedge *\omega(Y|x)
$$
The HS symmetry algebra is infinite dimensional and contains
$so[d,2]\oplus u(1)$. HS symmetry transformations satisfy the
composition rule
$$
{}[\varepsilon(Y|x)_{s_1},\varepsilon(Y|x)_{s_2}]_* =
\sum_{t=|s_1-s_2|+1}^{s_1+s_2-2}\varepsilon(Y|x)_{t}
$$
As soon as $s>2$ the algebra becomes infinite dimensional. The
Moyal-Weyl structure naturally leads to non-commutativity that in
turn can be related to the non-locality resulting from
interactions involving higher and higher derivatives. To some
extent a massless HS gauge theory is in between a field theory,
with spin ranging from 0 to 2, and string theory, where spin is
not bounded. Strings in AdS are the natural arena where to put
this analogy at work.


In the MDMSW description \cite{Vasiliev:2004qz} one can treat the Weyl tensor 
$C^{AC,BD}$ as an independent field and deform the field equations according
to
$$
R^{A,B}|_{oms} = E_C\wedge E_D C^{AC,BD}
$$
where $oms$ stands for `on mass-shell'. Similarly `unfolded' HS
dynamics can be defined in terms of generalized Weyl tensors
$C^{A_1..A_s,B_1...B_s}$ (two-row Young tableaux), that satisfy
the first on-mass-shell (oms) theorem
$$
R^{A_1..A_{s-1},B_1...B_{s-1}}|_{oms} = E_{A_s}\wedge E_{B_s} \
C^{A_1..A_{s},B_1...B_{s}}
$$
Linearizing around a maximally symmetric background yields
$$
R_1^{A_1...A_{s-1},B_1...B_{s-1}} = E_{0A_s}\wedge E_{0B_s}
C^{A_1...A_s,B_1...B_s}
$$
The full set of compatibility conditions on the basis
$C^{A_1...A_u,B_1...B_s}$ of $u-s$ derivatives of
$C^{A_1...A_s,B_1...B_s}$ is subsumed by
$$
\widetilde{D}_0 C^{A_1...A_u,B_1...B_s} = 0 \quad , \quad u\ge s
$$
where $$ \widetilde{D}_0={D}_{0L} - \Lambda E_0^A V^B(2
Y^{i\perp}_A Y^{\parallel}_{Bi} + {1\over 2} \varepsilon^{ij}
\partial^\perp_{iA}\partial^\parallel_{jB})
$$
is the background covariant derivative in the so-called `{\it
twisted} adjoint representation', where
$\partial_{iA}=\partial/\partial Y^{iA}$ and $X^{A\parallel} = V^A
V_B X^B$ and $X^{A\perp}= X^{A} - X^{A\parallel}$. 

At the linearized level, the `central on-mass-shell theorem'
implies
$$
R_1(Y^\parallel, Y^\perp)|x) = {1\over 2} E_0^A \wedge E_0^B
\varepsilon^{ij} \partial_{iA} \partial_{jB} C(0, Y^\perp |x)
$$
and
$$
\widetilde{D}_0 C(Y|x) = 0
$$
where $\widetilde{D}_0 C \equiv d C + \omega_0 *C -
C*\widetilde{\omega}_0$ and $R_1 \equiv d \omega + \omega_0 \wedge
*\omega + \omega\wedge *\widetilde{\omega}_0$

In Vasiliev's formulation \cite{Vasiliev:2004qz} of non-linear HS dynamics in 
$D=4$, one doubles the number of oscillators, adding $Z^A_i$ with
$[Z^A_i,Z^B_j]_* = - \eta^{AB}\epsilon_{ij}$, and works with three
`master fields': the connection one-form $W(Z,Y|x)= dx^n
W_n(Z,Y|x)$ such that $W(0,Y|x)=\omega(Y|x)$, the 0-form (scalar)
$B(Z,Y|x)$ such that $B(0,Y|x)=C(Y|x)$, and an `auxiliary'
one-form $S(Z,Y|x)= dZ^A_i S^i_A(Z,Y|x)$. One then generalizes the
Moyal-Weyl star product
$$
(f*g)(Z,Y) = \int {dT dS \over\pi^{2(d+2}} f(Z+S, Y+S) g(Z-T, Y+T)
\exp(-2S\cdot T)
$$
and defines the `inner Klein operator'
$$
{\cal K} = \exp(-2z_iy^i)
$$
with $x_i= V_A X^A_i/\sqrt{V\cdot V}$ such that
$$
{\cal K} * f = \tilde{f} * {\cal K} \quad , \quad {\cal K}*{\cal
K} = 1
$$
where $\tilde{f}(Z,Y) = f(\tilde{Z},\tilde{Y})$ with $\tilde{X}^A
= X^{A\perp}-X^{A\parallel}$. The full non-linear system of
Vasiliev's HS equations then reads
$$
dW + W * W = 0 \quad , \quad dS + W * S + S * W = 0 \quad , \quad
dB + W * B - B
* \tilde{W} = 0
$$
and
$$
S * S = - {1\over 2} (dZ^A_i dZ^i_A + {4\over \Lambda}dz_i dz^i B
* {\cal K}) \quad , \quad  S * B = B * \tilde{S}
$$
In terms of ${\cal W} = d + W + S$ one can combine both sets of
equations into
$$
{\cal W} * {\cal W} = - {1\over 2} (dZ^A_i dZ^i_A +
{4\over\Lambda}dz_i dz^i B * {\cal K}) \quad , \quad {\cal W} * B
= B * \tilde{\cal W}
$$
Formal consistency follows from associativity, while gauge
invariance under
$$
\delta{\cal W}=[\varepsilon,{\cal W}]_* \quad , \quad \delta B
=\varepsilon
* B - B * \tilde{\varepsilon}
$$
is manifest. It is remarkable that the fully non-linear dynamics
is presented in the form of a zero-curvature condition that leads to an 
integrable Cartan system.

The linearized HS field equations resulting from the positions
$$
W = W_0 + W_1 \quad , \quad S = S_0 + S_1 \quad , \quad B = B_0 +
B_1
$$
with
$$
W_0 = {1\over 2} \omega_0^{AB}Y^i_AY_{iB} \quad , \quad B_0 = 0
\quad S_0 = dZ^i_A Z_{iA}
$$
where $\omega_0$ has zero $SO(d,2)$ curvature so as to describe
$AdS_{d+1}$, are equivalent to Fronsdal's component HS field
equations in an AdS background.


Vasiliev's equations \cite{Vasiliev:2004qz} describe HS gauge fields in $D=4$ 
where only totally symmetric tensors are relevant. On the one hand one would
like to extend his analysis to the case of $HS(2,2|4)$ relevant to
${\cal N}=4$ SYM in $d=4$, $D=5$ bulk. Sezgin and Sundell \cite{Sezgin:2001yf, 
Sezgin:2002rt, Sezgin:2001zs} have
been able to write down field equations for the `massless'
$HS(2,2|4)$ doubleton ($L=2$). As in Vasiliev's case the field
content can be assembled into a master connection $A$ and a master
scalar (curvature) $\Phi$. The former transform in the adjoint
representation of $hs(2,2|4)$ and contains physical gauge fields
with $s\ge 1$ and $B=0,\pm 1$. The latter transform in the twisted
adjoint representation and contributes physical fields with spin
$s\le 1/2$ or $s\ge 1$ but $|B|\ge 3/2$ (self-dual two-form
potentials). The field strengths \be F_A = dA + A\wedge *A \quad ,
\quad D_A \Phi = d\Phi + A * \Phi -\Phi*\tilde{A} \ee transform
covariantly \be \delta F_A = [F_A,\epsilon]_*\quad , \quad \delta
D_A\Phi = D_A\Phi * \tilde\epsilon - \epsilon *D_A\Phi \ee under
gauge transformations \be \delta A = d \epsilon +
[A,\epsilon]_*\quad , \quad \delta \Phi = \Phi * \tilde\epsilon -
\epsilon *\Phi \ee The linearized constraints and integrability
conditions lead after some tedious algebra
to the correct linearized field equations for the `matter' fields
with $s\le 1/2$, for the HS gauge fields and for the antisymmetric
tensors with generalized self-duality.

Possibly because of the presence of these generalized (anti)self
dual tensors, an interacting $hs(2,2|4)$ gauge theory has not yet
been formulated. For the purpose of describing the breaking of
$hs(2,2|4)$ to $psu(2,2|4)$, however, one is rather interested in
the coupling of the `massive' HS multiplets (totally antisymmetric
`tripleton' and window-like `tetrapleton') containing the
Goldstone lower spin modes to the massless doubleton at large $N$.
This problem might turn out to be easier to solve than
constructing a fully non-linear massless $hs(2,2|4)$ theory
because it should be fixed by linearized HS symmetry and require a
little bit more than the knowledge of the linearized field
equations. In some sense equations of Vasiliev's type should
encode combinatorial interactions which are present even in a free
field theory at finite $N$, where HS symmetry is unbroken, or
couplings to multi-particle states\footnote{Precisely for this
reason they are relevant in the $d=3$ model on the boundary of
$AdS_4$ considered by Klebanov and Polyakov, for a recent review see \eg 
\cite{PetkouRTN}.}. 
From a holographic
perspective the interactions which are responsible for the
breaking of HS symmetry are equivalent to making the curvature
radius larger than the string scale. Although truncation to the HS
massless multiplet (doubleton) should be consistent at the point
of HS enhancement this should no more be the case for generic $R$ in $AdS_5$.

When interactions are turned on only
one out of the infinite tower of conserved current doubleton
multiplets in ${\cal N}=4$ SYM theory 
\[
\label{tksdjf} {\cal Z}_{\Yboxdim4pt \yng(2)}=\sum_{n=0}^\infty
\mult_{2n}, \qquad \mult_{j}:=\mult^{j,0
}_{[-1+\frac{1}{2}j^\ast,-1+\frac{1}{2}j^\ast][0,0,0]}.\]
is protected against quantum corrections to the scaling dimension:
the $\superN=4$ supercurrent multiplet
$\mult_0=\mult^{2,0}_{[0^\dagger,0^\dagger][0^\dagger,2,0^\dagger]}$.
The remaining multiplets $\mult_{2n}$ acquire anomalous dimensions
dual to mass shifts in the bulk which violate the conservation of
their HS currents at the quantum level. At one-loop, one has
\cite{Kotikov:2000pm, Kotikov:2001sc, Dolan:2001tt}
\[
\gamma_{\rm 1-loop}(2n) =\frac{\gym^2 N}{2\pi^2}\,h(2n),\qquad
h(j)=\sum_{k=1}^j\frac{1}{k},
\]
This elegant (`number theoretic') formula gives a clue on how to
compute generic anomalous dimensions at first order in
perturbation theory relying on symmetry breaking considerations.
Naively, one would look for all occurrences of the broken currents
$\mult_{2n}$ within some operator $\mathcal{O}$. Each occurrence
of some broken current should contribute to the anomalous
dimension of $\mathcal{O}$ a term proportional to $h(2n)$. Indeed,
this is nearly what happens, the one-loop dilatation operator
\cite{Beisert:2003jj} can be written as
\[
H=\sum_{s=1}^L H_{(s,s+1)} = \sum_{s=1}^L \sum_{j=0}^\infty 2h(j)
\,P^{j}_{(s,s+1)} ,
\]
where $P^{j}_{(s,s+1)}$ projects the product of fields (`letters')
at nearest neighboring sites $s$ and $s+1$ onto $\mult_{j}$. Here,
the sum goes over all values of $j$ and not just the even ones.
The point is that although bilinear currents $\mult_{2n+1}$
corresponding to the broken generators are eliminated after
tracing over color indices, they still  appear in subdiagrams
\,${\Yboxdim5pt \yng(1,1)}$\, inside a bigger trace.

In order to achieve a holographic description of {\it La Grande Bouffe}, 
our previous identification of the necessary `longitudinal' modes in the $AdS$
bulk, turns out to be crucial. Using the Konishi multiplet
as a prototype, one expects something like \be {\cal K}_{long}
\leftrightarrow {\cal K}_{short} + {\cal K}_{1/4} + {\cal K}_{1/8}
+ {\cal K}^*_{1/8} \ee \ie HS semishort multiplets, such as ${\cal
K}_{short}$, eat lower spin Goldstone multiplets, such as ${\cal K}_{1/8}$,
its conjugate ${\cal K}^*_{1/8}$ and ${\cal K}_{1/4}$. Although
massless HS fields with mixed symmetry have been only recently
addressed \cite{deMedeiros:2003dc}, whenever they are part of the HS doubleton 
multiplet,
supersymmetry should be enough to determine their equations from
the more familiar equations for symmetric tensors. In particular
for the ${\cal N}=4$ Konishi multiplet we know the axial anomaly
is part of an on-shell anomaly supermultiplet
\be \bar{D}^A \bar{D}^B {\cal K}_{long} = g_{ym} Tr({\cal
W}_{EF}[{\cal W}^{AE} {\cal W}^{BF}]) + {g_{ym}^2\over 8\pi^2} D_E
D_F Tr({\cal W}^{AE} {\cal W}^{BF}) \ee

The formulation of {\it La Grande Bouffe} we have in mind is of
the St\"uckelberg type \cite{Bianchi:2000sm, Bianchi:2001de, Bianchi:2001kw}.
Let us illustrate it for the broken
singlet current in the ${\cal N}=4$ Konishi multiplet. The bulk
Lagrangian describing the holographic Higgs mechanism \`a la
St\"uckelberg should schematically be of the form \be L = -
{1\over 4} F(V)^2 + {1\over 2} (\partial \alpha - M V)^2 \ee where
$F_{mn}$ is the field-strength of the  bulk vector field $V_{n}$ dual to
the current $J_\mu$ and $\alpha$ is the bulk scalar dual to
the `anomaly' $A =
\partial_\mu J^\mu$. Gauge invariance under \be \delta V_m =
\partial_m \vartheta \quad , \quad \delta \alpha = M  \vartheta\ee
is manifest for constant $M$. For $M=0$, $V$ and $\alpha$
decouple. For $M\neq 0$, $V$ eats $\alpha$ and becomes massive. In
practice $M$ should depend on the dilaton and the other massless
scalars. Since we want to preserve superconformal invariance, $M$
can at most acquire a constant vev and the above analysis for a vector current
seems robust. Generalization to higher spins and 
AdS covariantization should
not pose any fundamental problem. We will use string theory in flat space as a 
guidance in order to
write down gauge invariant field equations that allow for symmetry
breaking \`a la St\"uckelberg.


Let us start from the field equations for a massive HS
field of spin $s$ corresponding to a totally symmetric traceless tensor, \ie
$\phi_{(s)}$ with $\hat\phi_{(s-2)}=0$. Suppressing indices as in 
\cite{Riccioni, Bouatta:2004kk}, one
has \be
\partial^2_{(0)} \phi_{(s)} + M^2 \phi_{(s)} = 0
\quad , \quad \partial_{(-1)}\phi_{(s)}=0 \quad .\ee  

For spin $s$, the minimal set of auxiliary fields
identified by Singh and Hagen \cite{Singh:1974qz,Singh:1974rc} consists in 
$\phi_{(s-t)}\approx
\partial_{(-1)}^t\phi_{(s)}$ with $t=2,...s$. The Singh - Hagen
Lagrangian reads
\bea &&(-)^sL_{(s)}={1\over 2} (\partial_{(1)}\phi_{(s)})^2
-{s\over 2} (\partial_{(-1)}\phi_{(s)})^2 - {1\over 2} M^2
\phi_{(s)}^2 \\
&&+ C_s \{a_2 M^2 \phi_{(s-2)}^2 - {1\over 2}
(\partial_{(1)}\phi_{(s-2)})^2 +
\phi_{(s-2)}\partial_{(-1)}^2\phi_{(s)}+{b_2 \over
2}(\partial_{(-1)}\phi_{(s-2)})^2 \nonumber \\
&&- \sum_{t=3}^s \prod_{r=2}^{t-1} c_r [{1\over 2}
(\partial_{(1)}\phi_{(s-t)})^2- a_t M^2 \phi_{(s-t)}^2- {b_t \over
2}(\partial_{(-1)}\phi_{(s-t)})^2 - M
\phi_{(s-t)}\partial_{(-1)}\phi_{(s-t+1)}]\}\nonumber \eea where
$c_t=-{(t-1)(s-t)^2(s-t+2)(2s-t+2)/
[2(s-t+1)(2s-2t+1)(2s-2t+3)]}$, $C_s={s(s-1)^2/(2s-1)}$,
$a_t={t(2s-t+1)(s-t+2)/[2(2s-2t+3)(s-t+1)}]$, and $b_t=-{(s-t)^2/
(2s-2t+3)}$.

In the massless limit, $M\rightarrow 0$, HS equations enjoy gauge
invariance. The spin 1 case is very well known. For spin 2, the
relevant gauge transformations read $\delta\phi_{(2)} =
\partial_{(1)}\epsilon_{(1)} - {\eta_{(2)}\over 2}
\partial_{(-1)}\epsilon_{(1)}$, so as to preserve tracelessness, and
$\delta\phi_{(0)} = - {3 \over 2}\partial_{(-1)}\epsilon_{(1)}$.
It is convenient to introduce a new traceful field $h_{(2)} =
\phi_{(2)} - \eta_{(2)}\phi_{(0)}/3$, for which $\delta h_{(2)} =
\partial_{(1)}\epsilon_{(1)}$. The `new' field equations are
nothing but linearized Einstein equations \be
\partial^2_{(0)} h_{(2)} - 2 \partial_{(1)}
\partial_{(-1)}h_{(2)}+ \partial_{(1)}^2 \hat{h}_{(0)} = 0 \quad .\ee

For massless spin $s$, Fronsdal's field equations \be
\partial^2_{(0)} h_{(s)} - s \partial_{(1)}
(\partial_{(-1)}h_{(s)})+ {s (s-1)\over 2}\partial_{(1)}^2
\hat{h}_{(s-2)} = 0\ee with $\hat{\hat{h}}_{(s-4)}= 0$ are
invariant under $\delta h_{(s)} =
\partial_{(1)}\epsilon_{(s-1)}$ with $\hat{\epsilon}_{(s-3)}= 0$. For
$M=0$,
$\phi_{(s)}$ and $\phi_{(s-2)}$ in Singh - Hagen description
decouple from the rest. Introducing $h_{(s)}=\phi_{(s)}-
\eta_{(2)}\hat\phi_{(s-2)}/(2s-1)$ one gets \bea &&(-)^s L_{(s)}
={1\over 2} (\partial_{(1)}h_{(s)})^2 -
{s\over 2} (\partial_{(-1)}h_{(s)})^2 \\
&&- {s(s-1)\over 2} (\partial_{(1)}\hat{h}_{(s-2)})^2 -
{s(s-1)(s-2)\over 8} (\partial_{(-1)}\hat{h}_{(s-2)})^2 -
{s(s-1)\over 2} \hat{h}_{(s-2)}\partial_{(-1)}^2 {h}_{(s)}
\nonumber \eea that coincides with (\ref{hsaction}) after
suppressing indices.

The set of  St\"uckelberg fields that participate in the
spontaneous breaking of HS symmetry can be elegantly derived by
formal KK reduction of the (quadratic) massless HS lagrangian from
$D+1$ dimensions. The {\it a priori} complex modes \be
h_{(s)}(x,y) =
\sum_{t,M} \psi_{(s-t),M}(x) \exp(i M y)\ee
 generate a bunch of
lower spin modes that are needed in order to reproduce the
correct number of d.o.f.'s $\nu_{M\neq 0} (d,s)$. It is easy to
check that $\nu_{M\neq 0} (d,s) = \nu_{M=0} (d,s) + \nu_{M\neq 0}
(d,s-1)$. By iteration, one eventually gets $\nu_{M\neq 0} (d,s) =
\sum_{t=0}^s \nu_{M=0} (d,t)$. After reduction, \ie integration
over $y$, one can assume that all the $\psi$'s are real for
simplicity. More explicitly, from a massless spin $s$ field
$\Phi_{(s)}$ in $D+1$ dimensions one gets `massless' fields
$\phi_{(s-t)}$ in $D$ dimensions \be \Phi_{(0)}
\quad\rightarrow\quad \{\phi_{(s-t)},\, t=0,... s\} \ee that
altogether provide the d.o.f.'s of a massive spin $s$ field when the
double tracelessness conditions \be \hat{\hat\Phi}_{(s)}=0
\quad\rightarrow\quad \{\hat{\hat\phi}_{(s-t-4)} + 2
{\hat\phi}_{(s-t-4)} + \phi_{(s-t)}=0 ,\, t=0,... s-4\} \ee are
taken into account. The resulting field equations \bea
&&\partial_{(0)}^2 \Phi_{(s)} - s ?
\partial_{(1)}(\partial_{(-1)}\Phi_{(s)}) +
\partial_{(1)}(\partial_{(1)}\hat\Phi_{(s-2)})= 0 \quad
\rightarrow \\
&& \partial_{(0)}^2 \Phi_{(s-t)} - {(s-1)(s-2)\over 2}M^2
\Phi_{(s-t)} + {s(s-1)\over 2}M^2 \hat\Phi_{(s-t+2-2)} \nonumber
\\
&& - (s-t)? \partial_{(1)}(\partial_{(-1)}\Phi_{(s-t)}) +
\partial_{(1)}(\partial_{(1)}\hat\Phi_{(s-t-2)})
+ \partial_{(1)}(\partial_{(1)}\Phi_{(s-t-2)}) \nonumber \\
&& - t M \partial_{(-1)}\Phi_{(s-t+1)} + t M
\partial_{(-1)}\hat\Phi_{(s-t+1)} + (1-t) M (\partial_{(1)}\Phi_{(s-t-1)})= 0
\eea are invariant by construction under gauge transformations \be
\delta \Phi_{(s)} =
\partial_{(1)} {\cal E}_{(s-1)} \quad\rightarrow\quad \{\delta\phi_{(s-t)} =
\partial_{(1)} \epsilon_{(s-t-1)} + t M \epsilon_{(s-t)} ,\,
t=0,... s\} \ee with \be \hat{\cal E}_{(s-3)}=0
\quad\rightarrow\quad \{\hat\epsilon_{(s-t-3)} +
\epsilon_{(s-t-3)} ,\, t=0,... s-1\} \ee that expose the role of
the lower spin fields in the Higgsing of the HS symmetry.

In string theory the situation is slightly different \cite{Riccioni, 
Bouatta:2004kk}. Given the difficulties with the quantization of the type IIB  
in $AdS_5$, let us take the open bosonic
string in flat space as a toy model. States $|\Phi\rangle$ of zero ghost 
number can be obtained by
acting with `positive' frequency modes on the tachyonic
groundstate $|\Omega\rangle = c_{+1} |0\rangle_{SL(2)}$,
annihilated by the `negative' frequency modes of the bosonic
coordinates ($\alpha_n^\mu |\Omega\rangle =0$ with $n>0$) and the
ghosts ($c_n|\Omega\rangle =0$ for $n>0$, $b_n|\Omega\rangle =0$
for $n\ge 0$). Expanding in levels, $|\Phi\rangle = \sum_{\ell}
|\Phi\rangle_\ell$, yields
\bea &&|\Phi\rangle_{\ell=0} = T |\Omega\rangle \\
&&|\Phi\rangle_{\ell=1}= [A_\mu \alpha^\mu_{-1} + \rho c_0
b_{-1}]|\Omega\rangle
\\
&&|\Phi\rangle_{\ell=2}= [H_{\mu\nu}
\alpha^\mu_{-1}\alpha^\nu_{-1} + B_\mu \alpha^\mu_{-2} + \psi
c_{-1} b_{-1} + \chi_\mu \alpha^\mu_{-1} c_0 b_{-1} + \eta c_0
b_{-2}]|\Omega\rangle \\
&&|\Phi\rangle_{\ell=3}= [S_{\mu\nu\rho}
\alpha^\mu_{-1}\alpha^\nu_{-1} \alpha^\rho_{-1} + U_{\mu\nu}
\alpha^\mu_{-2}\alpha^\nu_{-1} + V_\mu \alpha^\mu_{-3} \nonumber \\
&&\quad + \varphi_{\mu\nu} \alpha^\mu_{-1} \alpha^\nu_{-1} c_0
b_{-1} + \omega_{\mu} \alpha^\mu_{-2} c_0 b_{-1} + \sigma_{\mu}
\alpha^\mu_{-1} c_0 b_{-2} + \tau_{\mu} \alpha^\mu_{-1} c_{-1}
b_{-1} \nonumber \\
&&\quad +\zeta c_0 b_{-3} + \vartheta c_{-1} b_{-2} + \gamma
c_{-2} b_{-1}]|\Omega\rangle \eea for the first few levels. Field
equations $Q |\Phi\rangle = 0$, with $Q= c_0 (L_0 -1) + b_0 M +
Q'$, are invariant under gauge transformations $\delta
|\Phi\rangle = Q |\Lambda\rangle$, with parameter $
|\Lambda\rangle = \sum_{\ell} |\Lambda\rangle_\ell$ of the form
\bea &&|\Lambda\rangle_{\ell=0} = 0 \\
&&|\Lambda\rangle_{\ell=1} = [\lambda b_{-1}]|\Omega\rangle
\\
&&|\Lambda\rangle_{\ell=2} = [\varepsilon_{\mu} \alpha^\mu_{-1}
b_{-1} + \theta b_{-2}]\alpha^\mu_{-1} \\
&&|\Lambda\rangle_{\ell=3}
=[\epsilon_{\mu\nu}\alpha^\mu_{-1}\alpha^\nu_{-1} b_{-1} +
\kappa_\mu \alpha^\mu_{-1} b_{-2} + \xi_\mu \alpha^\mu_{-2} b_{-1}
+ \mu b_{-3} + \nu c_0 b_{-1} b_{-2}\nonumber \\
&&\quad +\beta b_{-1} b_{-2}]|\Omega\rangle \eea The last term
$|\Lambda\rangle^{red}_{\ell=3}= \beta b_{-1}
b_{-2}|\Omega\rangle$ accounts for the reducibility of the gauge
transformation of the antisymmetric tensor present at $\ell=3$.
Fields and gauge parameters involving $c_0$ appear algebraically
and can be eliminated \cite{Riccioni, Bouatta:2004kk}. 

In components, setting $\ap=1/2$, one finds the following
equations. \bea \ell=0: \qquad
\partial^2 T - 2 T = 0 \qquad \delta T = 0 \eea
which is the Klein-Gordon equation for the open string tachyon $T$.
\bea \ell=1: \qquad &&\partial^2 A_\mu -
\partial_\mu \rho = 0
\qquad \partial^\mu A_\mu - \rho = 0 \\
&& \delta A_\mu = \partial_\mu \lambda \qquad\qquad \delta \rho =
\partial^2 \lambda \eea
which gives the familiar Maxwell equations and gauge
transformations for $A_\mu$ after eliminating the `auxiliary' field $\rho$.
\bea \ell=2: \qquad &&(\partial^2 + 2) H_{\mu\nu} - (\partial_\mu
\chi_\nu + \partial_\nu \chi_\mu) = 0 \\
&& (\partial^2 + 2) B_{\mu} - 2\partial_\mu \eta = 0 \\
&& (\partial^2 + 2) \psi + 2\partial^\mu \chi_\mu + 2 \eta = 0 \\
&& \partial^\mu H_{\mu\nu} + B_{\nu} - \chi_\nu  = 0 \\
&& H^{\mu}{}_{\mu} + 2\partial^\mu B_\mu + \eta = 0\eea
Elimination of the `auxiliary' fields $\chi_\mu$ and $\eta$ yields
a triplet of equations for a massive spin 2 field that can be
recast in Singh - Hagen form after some field redefinition.

The same situation prevails for the massive spin 3 field at level
$\ell=3$ after elimination of the `auxiliary' fields
$\varphi_{\mu\nu}$, $\omega_{\mu}$, $\sigma_{\mu}$, and $\zeta$
and some residual field redefinition.

The emergence in the tensionless limit of triplets of massless
covariant field equations \be
\partial^2_{(0)} \phi_{(s)}=\partial_{(1)} \chi_{(s-1)} \quad ,
\quad
\partial_{(-1)} \phi_{(s)}= \chi_{(s-1)} + \partial_{(1)} \psi_{(s-2)} \quad ,
\quad
\partial^2_{(0)} \psi_{(s-2)}=\partial_{(-1)} \chi_{(s-1)}\ee
with unconstrained gauge invariance \be
\delta\phi_{(s)}=\partial_{(1)} \xi_{(s-1)} \quad , \quad \delta
\chi_{(s-1)} = \partial^2_{(0)} \xi_{(s-1)} \quad , \quad
 \delta\psi_{(s-2)}=\partial_{(-1)} \xi_{(s-1)}\ee
 for totally symmetric fields $\phi_{(s)}$, $\chi_{(s-1)}$,
$\psi_{(s-2)}$ in the first Regge trajectory has been 
recently discussed \cite{Sagnotti:2003qa} and 
reviewed in \cite{Bouatta:2004kk}.

AdS covariantization for massless HS fields, as recently discussed in 
\cite{Mikhailov:2002bp, Sagnotti:2003qa},
can be achieved by keeping simple gauge transformations \be
\delta\phi_{(s)}=\nabla_{(1)} \xi_{(s-1)} \quad , \quad
 \delta\psi_{(s-2)}=\nabla_{(-1)} \xi_{(s-1)}\ee
and constraint \be \chi_{(s-1)} = \nabla_{(-1)}\phi_{(s)} -
\nabla_{(+1)} \psi_{(s-2)} \quad .\ee
The price one has to pay is a more complicated gauge transformation
for $\chi_{(s-1)}$ \be \delta\chi_{(s-1)} =
\nabla_{(0)}^2\xi_{(s-1)} + {(s-1)(3-s-D)\over L^2} \xi_{(s-1)} +
{2\over L^2} g_{(2)}\hat\xi_{(s-3)} \ee where $g_{(2)}$ is the AdS
metric and field equations for $\phi_{(s)}$ and $\psi_{(s-2)}$
\bea &&\nabla^2_{(0)} \phi_{(s)}=\nabla_{(1)} \chi_{(s-1)} +
{1\over L^2}\{ 8 g_{(2)}\psi_{(s-2)} - 2 g_{(2)} \hat\phi_{(s-2)}
+
[(2-s)(3-s-D)-s] \phi_{(s)}\} \\
&&\nabla^2_{(0)} \psi_{(s-2)}=\nabla_{(-1)} \chi_{(s-1)}+ {1\over
L^2} \{ [s(D+s-2)+6] \psi_{(s-2)} - 2 g_{(2)} \hat\psi_{(s-4)}-4
\hat\phi_{(s-2)}\} \eea

Many if not all the ingredients for {La Grande Bouffe} are at hand, it only 
remains to put them together and cook them up.

\section*{Acknowledgements}
It is a pleasure for me to thank N. Beisert, J. F. Morales, and H. Samleben
for a very enjoyable collaboration and the organizers of and the participants 
in the RTN-EXT Workshop in Kolymbari. 
My special thanks go to Elias Kiritsis for providing 
an excellent environment and creating a stimulating atmosphere. Let me also 
take this chance to acknowledge long lasting
collaborations on closely related topics with Mike Green, Stefano Kovacs, 
Giancarlo Rossi, Yassen Stanev, Dan Freedman, and Kostas Skenderis.
The largely unoriginal discussion of HS fields is based on what I learnt in 
endless but not pointless discussions with Misha Vasiliev, Per Sundell, Ergin
Sezgin, Augusto Sagnotti, Tassos Petkou, Fabio Riccioni, and Dario Francia.
This work was supported in part by I.N.F.N., by the EC programs
HPRN-CT-2000-00122, HPRN-CT-2000-00131 and HPRN-CT-2000-00148, by
the INTAS contract 99-1-590, by the MURST-COFIN contract
2001-025492 and by the NATO contract PST.CLG.978785. These lecture notes were 
completed while I was visiting MIT within the INFN-MIT ``Bruno Rossi'' 
exchange program. The warm hospitality of the members of the CTP at MIT is 
kindly acknowledged.

\appendix
\section{${\cal N}=4$ shortening}
\label{sa1}

Here we collect some notation for representations of the ${\cal
N}=4$ superconformal
 algebra $PSU(2,2|4)$ and their shortenings.
We denote by $\mult^{\Delta,B}_{[j,\jb][q_1,p,q_2]}$ a generic
long multiplet of $\alPSU(2,2|4)$ with HWS of conformal dimension
$\Delta$ and hypercharge $B$ in the ${\cal
R}_{[j,\jb][q_1,p,q_2]}$ representation of $\alSU(2)\times
\alSU(2)\times \alSU(4)$. As usual, $[q_1,p,q_2]$ are Dynkin
labels of $\alSU(4)$ while $[j,\jb]$ denote $\alSU(2)\times
\alSU(2)$ spins. The representation content of
$\mult^{\Delta,B}_{[j,\jb][q_1,p,q_2]}$ under the bosonic
subalgebra $\mathfrak{su}(2)\times
\mathfrak{su}(2)\times\mathfrak{su}(4)$ may be found from
evaluating the tensor product~$\mult^{2,0,0}_{[0,0][0,0,0]}
\chi_{[j,\jb][q_1,p,q_2]}^{\Delta-2,B,P}$, with the long Konishi
multiplet $\mult^{2,0,0}_{[0,0][0,0,0]}$, or explicitly by using
the Racah-Speiser algorithm as
\begin{eqnarray} \mult^{\Delta,B}_{[j,\jb][q_1,p,q_2]}
 &=&
\sum_{\epsilon_{A\alpha},\bar{\epsilon}^A_{\dot{\alpha}}\in
\{0,1\}} {\chi}_{[j,\jb][q_1,p,q_2]+ \epsilon_{A\alpha} {\cal
Q}^A_{\alpha}
 + \bar{\epsilon}^A_{\dot{\alpha}}
 \bar{{\cal Q}}_{A\dot{\alpha}}}\;,
 \label{susy}
 \end{eqnarray}
where ${\chi}_{[j,\jb][q_1,p,q_2]}$ are character polynomials and
the sum runs over the $2^{16}$ combinations of the $16$
supersymmetry charges ${\cal Q}^A{}_{\alpha}$, $\bar{\cal
Q}_{A\dot{\alpha}}$, $A=1, \dots, 4,\alpha,\dot{\alpha}=1,2$. In
oscillator notation\footnote{Notice the flip of notations for the
conjugate charges with respect to~\cite{Bianchi:2003wx} and the
unconventional use of oscillators in the denominator to mean
annihilation operators.}
\begin{eqnarray}
&&{\cal Q}^1{}_{\alpha} = a^\dagger_{\alpha}\,{c_1}\equiv
\frac{a_{\alpha}}{c_1}=[\pm \ft12,0][1,0,0] \;, \quad \bar{\cal
Q}_{1\dot{\alpha}}=b^\dagger_{\dot{\alpha}}\,c^\dagger_1
\equiv b_{\dot{\alpha}}\,c_1=[0,\pm\ft12][-1,0,0]\nn\\
&&{\cal Q}^2{}_{\alpha} = a^\dagger_{\alpha}\,{c_2}\equiv
\frac{a_{\alpha}}{c_2}=[\pm \ft12,0][-1,1,0] \;,\quad \bar{\cal
Q}_{2\dot{\alpha}}=b^\dagger_{\dot{\alpha}}\,c^\dagger_2\equiv
b_{\dot{\alpha}}\,c_2=[0,\pm\ft12][1,-1,0]
\;,\nn\\
&&{\cal Q}^3{}_{\alpha} = a^\dagger_{\alpha} \,d^\dagger_1\equiv
a_{\alpha} \,d_1=[\pm \ft12,0][0,-1,1] \;,\quad \bar{\cal
Q}_{3\dot{\alpha}}= b^\dagger_{\dot{\alpha}}\,d_1\equiv
\frac{b_{\dot{\alpha}}}{d_1}=[0,\pm\ft12][0,1,-1]
\;,\nn\\
&&{\cal Q}^4{}_{\alpha} = a^\dagger_{\alpha} \,d^\dagger_2\equiv
a_{\alpha} \,d_2=[\pm\ft12,0][0,0,-1] \;,\quad \bar{\cal
Q}_{4\dot{\alpha}}=b^\dagger_{\dot{\alpha}}\,d_2\equiv
\frac{b_{\dot{\alpha}}}{d_2}=[0,\pm\ft12][0,0,1] \;. \label{qs}
\end{eqnarray}
where $a_\alpha$, $b_{\dot\alpha}$, respectively $c_r$,
$d_{\dot{r}}$ appear in the decomposition of $y^a$ under
$\alSU(2)\times \alSU(2)\subset \alSU(2,2)$ and respectively of
$\theta^A$ under $\alSU(2)\times \alSU(2)\subset \alSU(4)$. Every
${\cal Q}^A{}_{\alpha}$, $\bar{\cal Q}_{A\dot{\alpha}}$
 raises the conformal dimension by $\ft12$,
parity is left invariant, and hypercharge $B$ is lowered and
raised by $\ft12$ by each ${\cal Q}^A{}_{\alpha}$ and $\bar{\cal
Q}_{A\dot{\alpha}}$ respectively. In order to make sense out of
\eqref{susy} also for small values of $q_1, p, q_2, j, \jb$, we
note that the character polynomials with negative Dynkin labels
are to be defined according to
\begin{eqnarray} {\chi}_{[j,\jb][q_1,p,q_2]}&=&
-{\chi}_{[j,\jb][-q_1-2,p+ q_1+ 1,q_2]} ~=~
-{\chi}_{[j,\jb][q_1,p+ q_2+1,- q_2-2]} \non &=&
-{\chi}_{[j,\jb][q_1+p+1,-p-2,q_2+p+1]} \non &=&
-{\chi}_{[-j-1,\jb][q_1,p,q_2]} ~=~
-{\chi}_{[j,-\jb-1][q_1,p,q_2]}
 \;.
\la{negw} \end{eqnarray}
In particular, this implies that ${\chi}_{[j,\jb][q_1,p,q_2]}$ is
identically zero whenever any of the weights $q_1$, $p$, $q_2$
takes the value~$-1$ or one of the spins $j$, $\jb$ equals
$-\ft12$.

In $\superN=4$ SYM, there are two types of (chiral) shortening
conditions for particular values of the conformal dimension
$\Delta$: BPS (B) which may occur when at least one of the spins
is zero, and semi-short (C) ones. The corresponding multiplets are
constructed similar to the long ones \eqref{susy}, with the sum
running only over a restricted number of supersymmetries.
Specifically, the critical values of the conformal dimensions and
the restrictions on the sums in \eqref{susy} are given by
\<\label{eq:shortening}
\begin{tabular}{llll}
B$_L$: \quad&
$\mult^{\Delta,B}_{[0^\dagger,\jb][q_1,p,q_2]}$\qquad\qquad &
$\Delta= p+\sfrac{3}{2}q_1+\sfrac{1}{2}q_2$ & $\epsilon_{1\pm}=0$
\\[1ex]
B$_R$: & $\mult^{\Delta,B}_{[j,0^\dagger][q_1,p,q_2]}$& $\Delta=
p+\sfrac{1}{2}q_1+\sfrac{3}{2}q_2$ & $\bar{\epsilon}_{4\pm}=0$
\\[1ex]
C$_L$: & $\mult^{\Delta,B}_{[j^*,\jb][q_1,p,q_2]}$&
$\Delta=2+2j+p+\sfrac{3}{2}q_1+\sfrac{1}{2}q_2$ &
$\epsilon_{1-}=0$
\\[1ex]
C$_R$: & $\mult^{\Delta,B}_{[j,\jb^*][q_1,p,q_2]}$ &
$\Delta=2+2\jb+p+\sfrac{1}{2}q_1+\sfrac{3}{2}q_2$ \qquad\qquad&
$\bar{\epsilon}_{4-}=0$
 \end{tabular}
 \>
for the different types of multiplets. They represent the basic
$\ft18$-BPS and $\ft{1}{16}$ semishortenings in ${\cal N}=4$ SCA
and are indicated as in with a ``$\dagger$'' and a ``$*$''
respectively.

If the conformal dimension $\Delta$ of the HWS of a long multiplet
satisfies one of the conditions~\eqref{eq:shortening},
the multiplet splits according to
\<\label{eq:splitting0} \mathrm{L}:\quad
\mult^{\Delta,B}_{[j,\jb][q_1,p,q_2]}\eq
\mult^{\Delta,B}_{[j^\ast,\jb][q_1,p,q_2]}+ \mult^{\Delta+
\frac{1}{2},B-\frac{1}{2}}_{[j-
\frac{1}{2}^\ast,\jb][q_1+1,p,q_2]}\;, \quad \nln \mathrm{R}:\quad
\mult^{\Delta,B}_{[j,\jb][q_1,p,q_2]}\eq
\mult^{\Delta,B}_{[j,\jb^\ast][q_1,p,q_2]}+
\mult^{\Delta+\frac{1}{2},B+ \frac{1}{2}}_{[j,\jb-
\frac{1}{2}^\ast][q_1,p,q_2+1]}\;, \>
where by `$^*$' we denote the $1/16$ semishortening. Consequently,
we denote by $\mult^{\Delta,B}_{[j^\ast,\jb^\ast][q_1,p,q_2]}$ the
$1/8$ semi-short multiplets appearing in the decomposition
\begin{eqnarray} \mult^{\Delta,B}_{[j,\jb][q_1,p,q_2]}\eq
\mult^{\Delta,B}_{[j^\ast,\jb^\ast][q_1,p,q_2]}+ \mult^{\Delta+
\frac{1}{2},B-\frac{1}{2}}_{[j-
\frac{1}{2}^\ast,\jb^\ast][q_1+1,p,q_2]}+
\mult^{\Delta+\frac{1}{2},B+ \frac{1}{2}}_{[j^\ast,\jb-
\frac{1}{2}^\ast][q_1,p,q_2+1]}\non &&{} +
\mult^{\Delta+1,B}_{[j-\frac12^\ast,\jb-
\frac{1}{2}^\ast][q_1+1,p,q_2+1]} \;,
\label{ltoshort}\end{eqnarray}
if left and right shortening conditions in \eqref{eq:shortening}
are simultaneously satisfied. The semishort multiplets appearing
in this decomposition are constructed explicitly according to
\eqref{susy}, \eqref{eq:shortening}.

Formulae \eqref{eq:splitting0} include the special cases
$\mult^{\Delta,B}_{[j^\ast,\jb][0,p,q_2]}$,
$\mult^{\Delta,B}_{[j^\ast,\jb][0,0,q_2]}$, and
$\mult^{\Delta,B}_{[j^\ast,\jb][0,0,0]}$, corresponding to
(chiral) $1/8$, $3/16$, and $1/4$ semi-shortening, respectively;
likewise for $\mult^{\Delta,B}_{[j,\jb^\ast][q_1,p,0]}$,
$\mult^{\Delta,B}_{[j,\jb^\ast][q_1,0,0]}$, and
$\mult^{\Delta,B}_{[j,\jb^\ast][0,0,0]}$. For $j=0$ and $\jb=0$,
respectively, the decompositions~\eqref{eq:splitting0} yield
negative spin labels. They are to be interpreted as BPS
multiplets, denoted by `$^\dagger$', as follows
\begin{eqnarray}
\mult^{\Delta,B}_{[- \frac{1}{2}^\ast,\jb][q_1,p,q_2]} \equiv
\mult^{\Delta+ \frac{1}{ 2},B+
\frac{1}{2}}_{[0^\dagger,\jb][q_1+1,p,q_2]}\;, \qquad
\mult^{\Delta,B}_{[j,- \frac{1}{2}^\ast][q_1,p,q_2]} \equiv
\mult^{\Delta+ \frac{1}{2},B-
\frac{1}{2}}_{[j,0^\dagger][q_1,p,q_2+1]}\label{bps} \;,
\end{eqnarray}
where one verifies that the BPS highest weight states satisfy the
BPS shortening conditions of \eqref{eq:shortening}. In addition,
there is the series
$\mult^{p,0}_{[0^{\dagger},0^{\dagger}][0^\dagger,p,0^\dagger]}$
of $\ft12$-BPS multiplets.

For convenience (but not quite accurately) we can also define
\[
\mult^{p,0}_{[-1^\ast,-1^\ast][0,p,0]} :=
\mult^{p+2,0}_{[0^{\dagger},0^{\dagger}][0^\dagger,p+2,0^\dagger]}.
\label{bpsnot}
\]


\end{document}